\documentclass[conference]{IEEEtran}
\IEEEoverridecommandlockouts
\usepackage{cite}
\usepackage{amsmath,amssymb,amsfonts}
\usepackage{algorithmic}
\usepackage{booktabs}
\usepackage{graphicx}
\usepackage{textcomp}
\usepackage{adjustbox}
\usepackage{subcaption}
\usepackage{listings}
\usepackage{caption}
\usepackage{color}
\usepackage[bookmarks=false]{hyperref}
\def\fixme#1{\typeout{FIXED in page \thepage : {#1}}
\bgroup \color{red}{[FIXME: {#1}]} \egroup}

\def\BibTeX{{\rm B\kern-.05em{\sc i\kern-.025em b}\kern-.08em
    T\kern-.1667em\lower.7ex\hbox{E}\kern-.125emX}}
\begin{document}

\title{Denial-of-Service Attacks on Shared Cache in Multicore: Analysis and Prevention}

\pagestyle{empty}
 \author{Michael G. Bechtel, Heechul Yun\\
   University of Kansas, USA. \\
   \{mbechtel, heechul.yun\}@ku.edu \\
 }

\maketitle

\begin{abstract}
  In this paper we investigate the feasibility of denial-of-service
  (DoS) attacks on shared caches in multicore platforms. With carefully
  engineered attacker tasks, we are able to cause more than 300X
  execution time increases on a victim task running on a dedicated
  core on a popular embedded multicore platform, regardless of whether
  we partition its shared cache or not.
  Based on careful experimentation on real and simulated multicore
  platforms,
  we identify an internal hardware structure of a non-blocking cache,
  namely the cache writeback buffer, as a potential target of shared cache
  DoS attacks.
  We propose an OS-level solution to prevent such DoS attacks by extending
  a state-of-the-art memory bandwidth regulation mechanism. We implement
  the proposed mechanism in Linux on a real multicore platform and
  show its effectiveness in protecting against cache DoS attacks.
\end{abstract}


\section{Introduction} \label{sec:intro}

Efficient multi-core computing platforms are increasingly demanded in
embedded real-time systems, such as cars, to improve their
intelligence while meeting cost, size, weight and power constraints.
However, because many tasks in such a system may have strict
real-time requirements, unpredictable timing of multicore platforms
due to contention in the shared memory hierarchy is a major
challenge~\cite{bosch2019challenge}.
Moreover, timing variability in multicore platforms
can potentially be exploited by attackers.

For example, as cars are becoming more connected, there are increasing
possibilities for an attacker to execute malicious programs inside of a car's
computer~\cite{checkoway2011comprehensive}, such as via downloaded
3rd party apps. Even if the computer's runtime (OS and hypervisor)
strictly partitions cores (and memory) to isolate these potentially
dangerous programs from critical tasks, as long as they
share the same multicore computing platform, an attacker may still be
able to cause  significant timing influence to the critical
tasks---simply by accessing shared hardware resources, such as a
shared cache, at a high-speed, effectively mounting
\emph{denial-of-service (DoS) attacks}.

On modern multicore processors, non-blocking
caches~\cite{kroft1981lockup} are commonly used,
especially as shared last-level caches, to support
concurrent memory accesses from multiple cores. However, access to a
non-blocking cache can be blocked when any of the
cache's internal hardware buffers become full, after which the cache will
deny any further requests from the CPU---even if the request is
actually a cache-hit---until the internal buffers become available
again~\cite{gem5memory,shen2013modern}.
On a shared cache, such blocking globally affects \emph{all}
cores because none of the cores can access the cache until it
is unblocked, which can take a long time (e.g., hundreds of CPU cycles)
as it may need to access slow main memory. Therefore, if an attacker
can intentionally induce shared cache blocking, it can cause
significant timing impacts to other tasks on different cores. This is
because every shared cache hit access, which would normally take a few
cycles, would instead take a few hundreds cycles until the cache is
unblocked.

While most prior works on cache isolation focused on cache space
partitioning, either in software or in
hardware~\cite{kim2017attacking,ward2013ecrts,kim2013coordinated,mancuso2013rtas,kessler1992page,wolfe1994software,liedtke1997controlled,suh2002new,kim2004fair,chandra2005predicting},
these cache partitioning techniques are ineffective in preventing such
shared cache blocking because,
even if the cache space is partitioned, the cache's internal buffers may still
be shared. A recent study~\cite{valsan2016taming} experimentally showed
the ineffectiveness of cache partitioning in preventing shared cache blocking on a
number of out-of-order multicore platforms. Specifically, when the
miss-status-holding-registers (MSHRs)~\cite{kroft1981lockup}---which
track outstanding cache-misses---of a shared cache are exhausted,
consequent cache blocking can cause substantial---reportedly up to
21X---task execution time increases, even when
the task runs alone on a dedicated core with a dedicated cache partition
and its working-set fits entirely in its cache partition~\cite{valsan2016taming}.

In this paper, we first experimentally investigate the feasibility and
severity of cache DoS attacks on shared caches on contemporary
embedded multicore platforms: five quad-core CPU implementations,
out of which four are in-order and one is out-of-order architecture.
Our first main finding is that extreme shared cache blocking can occur
not only in out-of-order processors, as suggested
in~\cite{valsan2016taming}, but also in \emph{in-order} processors. In fact,
we observe the most severe execution time
increase---\emph{up to 346X} (times, not percent)---on a popular
in-order architecture based quad-core platform (Raspberry
Pi 3). This is surprising because it was believed that simpler in-order
architectures are less susceptible to memory contention than
out-of-order architectures~\cite{valsan2016taming,ungerer2013parmerasa}.

We use a cycle-accurate full system simulator
(Gem5~\cite{binkert2011gem5} and Ramulator~\cite{kim2016ramulator} for
CPU and memory, respectively) to identify possible
causes of such severe timing impacts of cache DoS attacks under
various microarchitectural configurations. Our findings include: (1)
eliminating MSHR contention alone
is not sufficient to mitigate potential cache DoS attacks because
another cache internal hardware structure, namely the \emph{writeback
  buffer}~\cite{shen2013modern}---which is used to temporarily store
evicted cache-lines---in a shared cache can be
an important additional source of cache blocking; (2) the combination of
a small cache writeback buffer and the presence of aggressive
hardware prefetchers can cause severe writeback buffer contention, and
subsequent cache blocking.

We propose an OS-level solution to mitigate shared cache DoS attacks
that target cache writeback buffers. Our solution is
based on MemGuard~\cite{yun2013rtas}, which is a Linux kernel module
that regulates (throttles) each core's maximum memory bandwidth
usage to a certain threshold value at a regular interval (e.g., 1ms)
using a hardware performance counter.
Our extension is to apply two separate regulations---one for
read (cache-line refills) and one for write (cache write-backs) memory
traffic---for each core.
This allows us to set a high threshold value for read bandwidth
while setting a low threshold value for write bandwidth. This
mitigates writeback buffer DoS attacks with minimal performance
impacts for normal, non-attacker applications, which are typically more
sensitive to read performance~\cite{shen2013modern,hennessy2011computer}.


Our solution is implemented on a real multicore platform and evaluated
against cache DoS attack programs that generate very high write
traffic (to overflow shared cache writeback
buffers.) The results show that it is effective in preventing such
cache DoS attacks while minimizing the throughput loss of the prior
memory bandwidth throttling approach.



This paper makes the following \textbf{contributions}:
\begin{itemize}
  \item We experimentally demonstrate the feasibility of shared cache
    DoS attacks on a number of contemporary embedded multicore
    platforms. In particular, we show that
    even relatively simple in-order multicore architectures can also
    be highly affected by such microarchitectural attacks.

  \item We provide detailed microarchitectural analysis as to why these
    cache DoS attacks are effective in modern multicore architectures
    featuring non-blocking caches and hardware prefetchers.
    In particular, we identify the writeback buffer of a shared cache
    as a potential attack vector that enables shared cache DoS attacks.

  \item We propose an OS-level solution to mitigate DoS attacks
    targeting a shared cache's writeback
    buffer. 
    The proposed OS solution is shown to be effective
    in mitigating such attacks. We also provide it as open-source\footnote{https://github.com/mbechtel2/memguard.}.
\end{itemize}

The remainder of this paper is organized as follows:
Section~\ref{sec:background} provides necessary background information on
non-blocking caches and hardware prefetchers. Section~\ref{sec:attack}
defines the threat model.
Section~\ref{sec:evaluation} shows 
the feasibility of shared cache DoS attacks on contemporary embedded
multicore platforms. Section~\ref{sec:simulation} validates the problem
of writeback buffer induced shared cache blocking using a simulated
multicore platform. Section~\ref{sec:solutions} presents our software
solution to counter shared cache DoS attacks. We discuss related work in
Section~\ref{sec:related} and conclude in Section~\ref{sec:conclusion}.

\section{Background} \label{sec:background}

In this section, we provide background on non-blocking caches
and hardware prefetchers.

\subsection{Non-blocking Cache}

\begin{figure}[h]
  \centering
  \includegraphics[width=.45\textwidth]{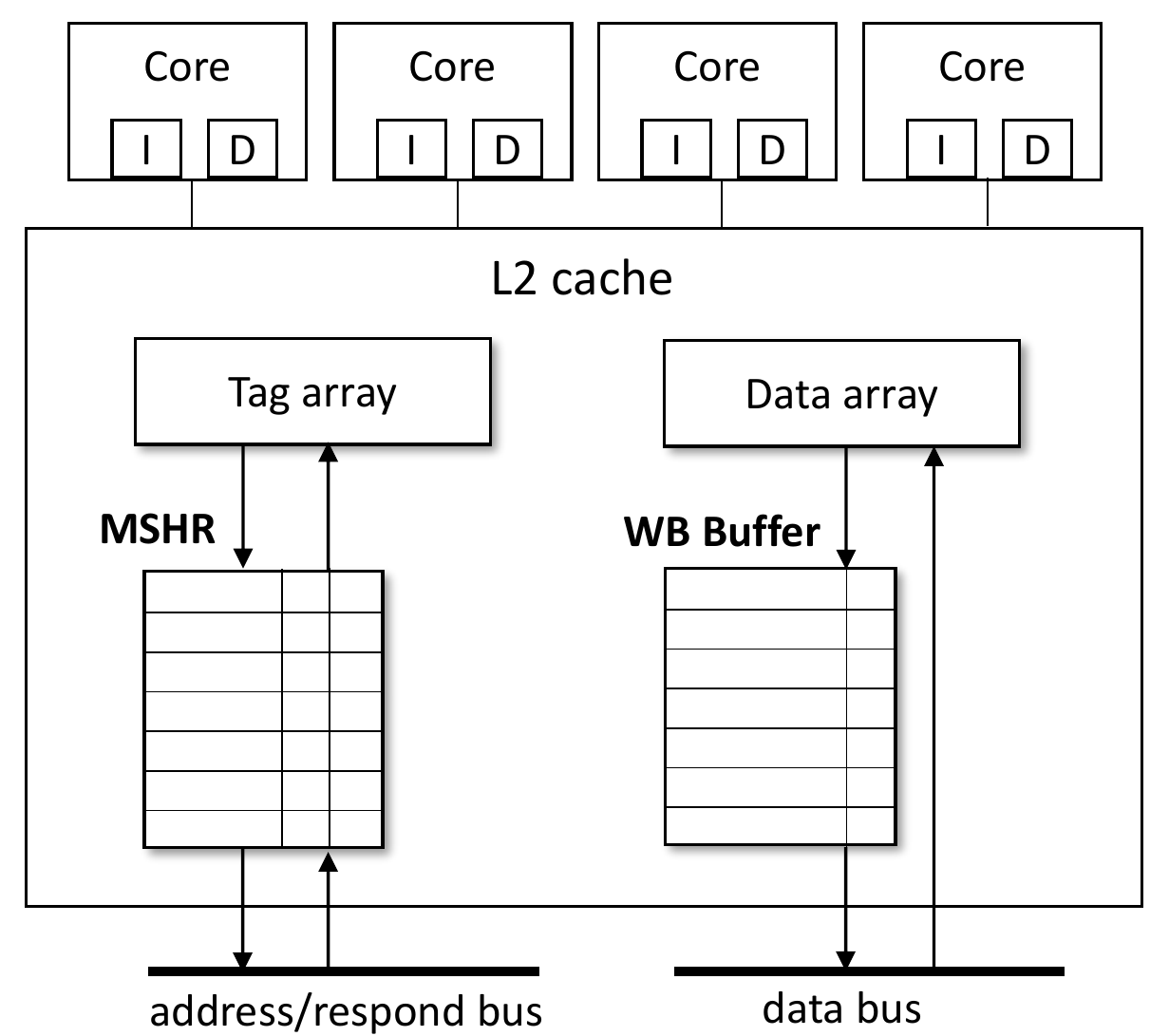}
  \caption{Internal organization of a shared L2 cache. Adopted from
    Figure 11.10 in~\cite{shen2013modern}.}
  \label{fig:l2cache}
\end{figure}

Modern processors employ non-blocking caches to improve cache-level
parallelism and performance. On a non-blocking cache, access to the
cache is allowed even while it is still processing prior cache
miss(es). This non-blocking access capability is essential in a multicore
processor, especially for the shared last-level cache, to achieve
high performance. Figure~\ref{fig:l2cache} shows the internal
structure of a non-blocking shared L2 cache of a multicore processor,
which depicts its two major hardware structures:
\emph{Miss-Status-Holding-Register (MSHR)} and \emph{WriteBack (WB) buffer.}

When a cache-miss occurs on a non-blocking cache, the miss related
information is recorded on a MSHR entry. The MSHR entry is cleared
once the corresponding cache-line is fetched from the lower-level
memory hierarchy (e.g., DRAM controller). In the mean time, the cache
can continue to serve concurrent memory requests from the CPU cores or
other higher-level caches (e.g., L1 caches). A non-blocking cache can
support multiple outstanding cache-misses, although the degree to
which this can occur depends on the size of the MSHR structure.

On the other hand, the writeback buffer is used to temporarily store
evicted (dirty) cache-lines from the cache while the corresponding
demand misses are serviced (cache-line refills). Because cache-line
refills (reads from DRAM) are generally more important to application
performance, delaying the writebacks reduces bus contention and,
therefore, improves performance.
In this way, a non-blocking cache can support concurrent access to the
cache efficiently most of the time.

Note, however, that when \emph{either} the MSHRs or writeback buffer of a
non-blocking cache becomes full, the entire cache is blocked---i.e., it
no longer accepts any further requests---until after free entries in
both the MSHRs and writeback buffer are available, at which point the cache
is said to be unblocked. Unfortunately, this can take a long time
because access to DRAM may take hundreds of CPU cycles, which could take
even longer if the DRAM controller is congested. When a shared cache
is blocked due to either MSHR or writeback buffer exhaustion, it
affects \emph{all} cores. If a task frequently accesses the
shared cache, even if the accesses are all cache-hits, the task may
still suffer massive execution time increase if the cache is blocked
most of the time.



\subsection{Hardware Prefetcher}

Cache prefetching is a technique used to reduce cache miss penalties by
preemptively loading memory blocks that are likely to be accessed in
the near future into the cache. Due to the high cost of a cache
miss, successful prefetching can significantly improve performance.
Therefore, modern processors often employ multiple hardware
prefetchers alongside the caches. A hardware prefetcher monitors
access to a cache and predicts future memory addresses based on
detected memory access patterns. It then generates a set number of requests
that are stored in its internal queue before being sent to the cache. This number
is called the prefetcher's \emph{degree} or \emph{depth}. Next-line and stride based prefetchers
are the most common.
However, while generally effective, prefetching can also incur unnecessary
cache-line refills due to mis-predictions, evict useful cache-lines,
pollute the cache, and generally add more pressure to the memory
hierarchy, which in turn can lead to increased cache blocking as we
will show in Section~\ref{sec:evaluation}.

\section{Threat Model}\label{sec:attack}

\begin{figure}[h]
  \centering
  \includegraphics[width=.38\textwidth,height=0.2\textheight]{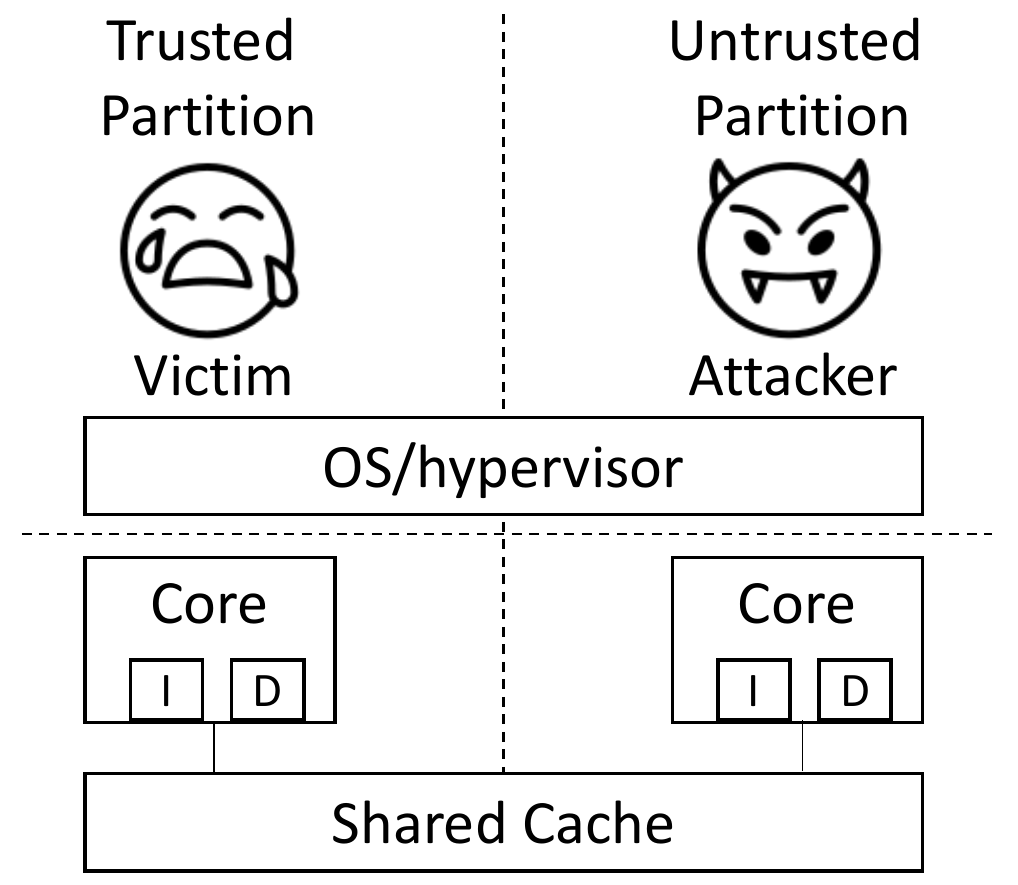}
  \caption[what]{Threat model.\footnotemark}
  \label{fig:threatmodel}
\end{figure}

We assume the victim and the attacker are co-located on a multicore
processor as shown in Figure~\ref{fig:threatmodel}. We assume the
multicore processor has a shared last-level cache as well as per-core
private caches. We assume that the runtime (OS and/or hypervisor)
provides core and memory isolation between the attacker and the
victim. In other words, the attacker cannot run on the same core as
the victim and cannot directly access the victim's memory. We henceforth
refer to the core that the victim runs on as the victim core, and the
core that the attacker runs on as the attacker core. In addition, we
assume that the runtime can partition cache space between the attacker
core and the victim core by means of page
coloring~\cite{kessler1992page,yun2014rtas}.

The only capability the attacker has is to run non-privileged program(s)
on the attacker core.
In this paper, the attacker's main goal is to generate timing
interference to the tasks running on the victim core. On a car, for
example, the attacker's goal may be to delay execution of real-time
control tasks running on a victim core so that they miss their deadlines,
potentially resulting in a crash. The attacker intends to achieve this goal by
mounting denial-of-service (DoS) attacks on shared hardware resources,
primarily the shared cache.

\footnotetext{The icons are by icons8: \url{https://icons8.com/}}

\section{Shared Cache DoS Attacks on Embedded Multicore Platforms}\label{sec:evaluation}
In this section, we experimentally evaluate the feasibility and
significance of cache denial-of-service (DoS) attacks on contemporary
embedded multicore platforms.


\subsection{Cache DoS Attack Code}

The main objective of the cache DoS attack code is to generate as many
concurrent cache-misses on the target cache as quickly as
possible. As discussed in Section~\ref{sec:background}, concurrent
cache-misses can exhaust available MSHR and writeback buffer
resources, and thereby induce cache blocking.


\begin{figure}[h]
  \centering
  \begin{subfigure}[b]{0.55\textwidth}
\begin{lstlisting}[linewidth=0.45\textwidth,language=c]
for (i = 0; i < mem_size; i += LINE_SIZE)
{
	sum += ptr[i];
}
\end{lstlisting}
    \caption{Read attack (BwRead)}
    \label{fig:read}
  \end{subfigure}
  \begin{subfigure}[b]{0.55\textwidth}
\begin{lstlisting}[linewidth=0.45\textwidth,language=c]
for (i = 0; i < mem_size; i += LINE_SIZE)
{
	ptr[i] = 0xff;
}
\end{lstlisting}
    \caption{Write attack (BwWrite)}
    \label{fig:write}
  \end{subfigure}
  \caption{Memory attacks. \texttt{LINE\_SIZE} = a cache-line size.}
  \label{fig:attackcode}
\end{figure}

Figure~\ref{fig:attackcode} shows read and write attack code
snippets. The read attack code simply iterates
over a single one-dimensional array, at every cache-line (LINE\_SIZE)
distance, and sums them up. Because \emph{sum} is allocated on a
register by the compiler, the code essentially keeps generating memory
\emph{load} operations, which may always miss the target cache if the
array size is bigger than the cache size. These missed loads will
stress the cache's MSHR.

\begin{table*}[h]
  \centering
  \begin{adjustbox}{width=0.85\textwidth}
  \begin{tabular}{|c|c|c|c|c|c|}
    \hline
    Platform   & Raspberry Pi 3    & Odroid C2   & Raspberry Pi 2 & \multicolumn{2}{|c|}{Odroid XU4} \\ \hline 
    SoC        & BCM2837           & AmlogicS905 & BCM2836        & \multicolumn{2}{|c|}{Exynos5422} \\ \hline 
    CPU     & 4x Cortex-A53 &
            4x Cortex-A53 &
	    4x Cortex-A7 &
	    4x Cortex-A7 &
            4x Cortex-A15\\
            & in-order
            & in-order
            & in-order
            & in-order
            & out-of-order\\
            & 1.2GHz
            & 1.5GHz
            & 900MHz
            & 1.4GHz
            & 2.0GHz\\
    \hline
    Private Cache & 32/32KB & 32/32KB & 32/32KB & 32/32KB & 32/32KB \\
    Shared Cache  & 512KB (16-way)& 512KB (16-way) & 256KB (8-way) & 512KB (16-way) & 2MB (16-way)\\
    \hline
    Memory      & 1GB LPDDR2	& 2GB DDR3	& 1GB LPDDR  &
    \multicolumn{2}{|c|}{2GB LPDDR3} \\ 
    (Peak BW)   & (8.5GB/s) 	& (12.8GB/s) 	& (8.5GB/s)  &
    \multicolumn{2}{|c|}{(14.9GB/s)} \\ 
	\hline
  \end{tabular}
  \end{adjustbox}
  \caption{Compared embedded multicore platforms.}
  \label{tbl:platforms}
\end{table*}

On the other hand, the write attack code performs the same
iteration over an array but, instead of reading each array entry, it
updates them, thereby generating memory \emph{store} operations. On a
writeback cache, each missed store will trigger two memory
transactions: one memory read (cache-line fill) and one memory write
(writeback of the evicted cache-line). Therefore, these missed stores
will stress both the MSHRs and writeback buffer of a cache.

We henceforth refer to the read attack as BwRead and the write attack
as BwWrite. In addition, their array sizes are denoted in
parentheses. For example, (LLC) denotes that the attacker's
working-set is configured  so that it always hits the shared
last-level cache but misses the private L1 cache (i.e., less than
1/4 of the LLC cache size but bigger than the L1 cache size). On the
other hand, (DRAM) denotes that the working-set size is bigger than
the shared LLC size.

\subsection{Embedded Multicore Platforms}

We evaluate the effectiveness of the attacks described above on four
embedded multicore platforms: Raspberry Pi 3, Odroid C2, Raspberry Pi
2, and Odroid-XU4. 
Both the Raspberry Pi 3 and the Odroid C2 employ four
Cortex A53 cores, while the Raspberry Pi 2 equips four Cortex A7 cores.
The Odroid XU4 has four Cortex-A7 and four Cortex-A15 cores in a
``big-little''~\cite{greenhalgh2011big} configuration.
Note that Cortex-A15 is a sophisticated out-of-order
design~\cite{arm-cortex-a15}, while Cortex-A7 and
Cortex-A53 are ``simpler'' in-order designs~\cite{arm-cortex-a7,
  arm-cortex-a53}.
In total, we compare five system configurations: four in-order and one
out-of-order designs. All platform specifications can be seen in
Table~\ref{tbl:platforms}.

\subsection{Synthetic Workloads}~\label{sec:eval-synthetic}

In this experiment, we use a BwRead (LLC) (Figure~\ref{fig:read})
as a synthetic victim task and evaluate the feasibility and severity of
shared cache DoS attacks.

The basic experiment setup is as follows: we first run the victim task
on Core0 and measure its solo execution time. We then co-schedule an
increasing number of attacker tasks on the other cores (Core1-3) and
measure the response times of the victim task. For attackers, we
use both BwRead (DRAM) and BwWrite (DRAM) to stress the L2 MSHR and
writeback buffer, respectively.


\begin{figure}[h]
  \centering
  \begin{subfigure}{0.45\textwidth}
    \includegraphics[width=\textwidth]{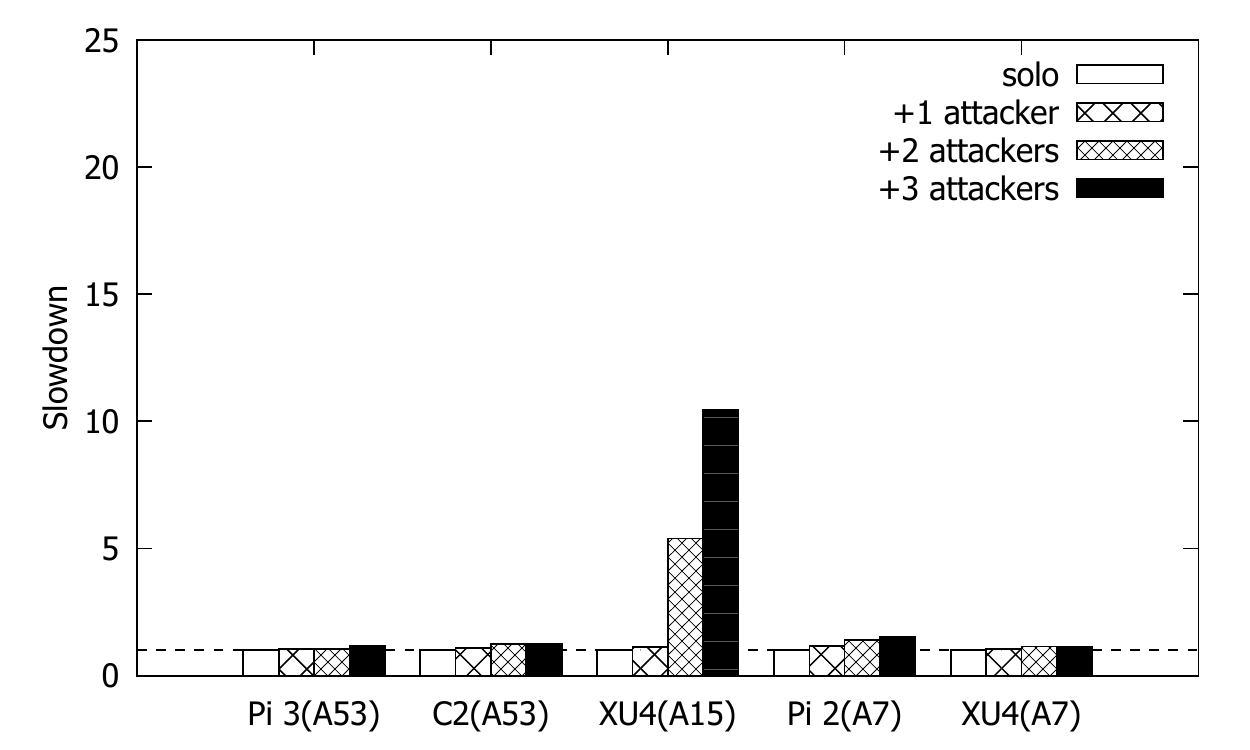}
    \caption{Effects of read attackers: BwRead(DRAM) }
    \label{fig:sys_bench_read}
  \end{subfigure}
  \begin{subfigure}{0.45\textwidth}
    \includegraphics[width=\textwidth]{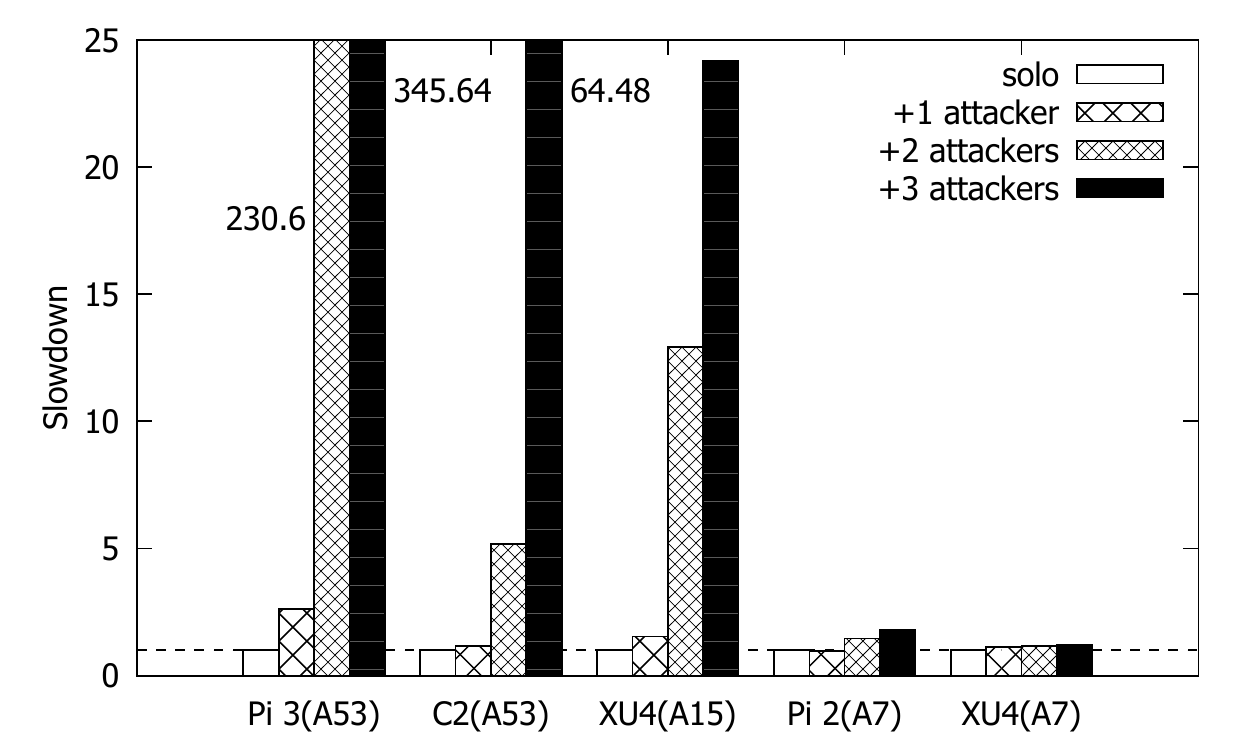}
    \caption{Effects of write attackers: BwWrite(DRAM)}
    \label{fig:sys_bench_write}
  \end{subfigure}
  \caption{Effects of memory attackers on a BwRead(LLC)
    victim. The attackers and victim run on their own dedicated cores.}
  \label{fig:compare-cache}
\end{figure}

Figure~\ref{fig:sys_bench_read} shows the victim task's performance
impact in the presence of BwRead (DRAM) attackers. Note first
that Odroid-XU4's Cortex-A15, which is an \emph{out-of-order}
architecture based CPU, suffers considerable execution time increases
from the read attackers, while the rest of the tested platforms,
the four \emph{in-order} architecture based ones,
show little performance impacts.
This result is consistent with the findings
in~\cite{valsan2016taming}, which suggested that an out-of-order core
can generate many concurrent cache accesses and when they all miss
the shared cache, due to the attacker's working set size being bigger than the cache,
they can cause cache blocking when all
MSHR entries are exhausted---i.e., MSHR contention.
In~\cite{valsan2016taming},
it is also suggested that \emph{in-order} cores are less likely to suffer
MSHR related cache blocking because an in-order core's ability to
generate concurrent memory accesses is limited---that is, one memory
access at a time.

\begin{figure}[t]
  \centering
  \begin{subfigure}{0.45\textwidth}
    \includegraphics[width=\textwidth]{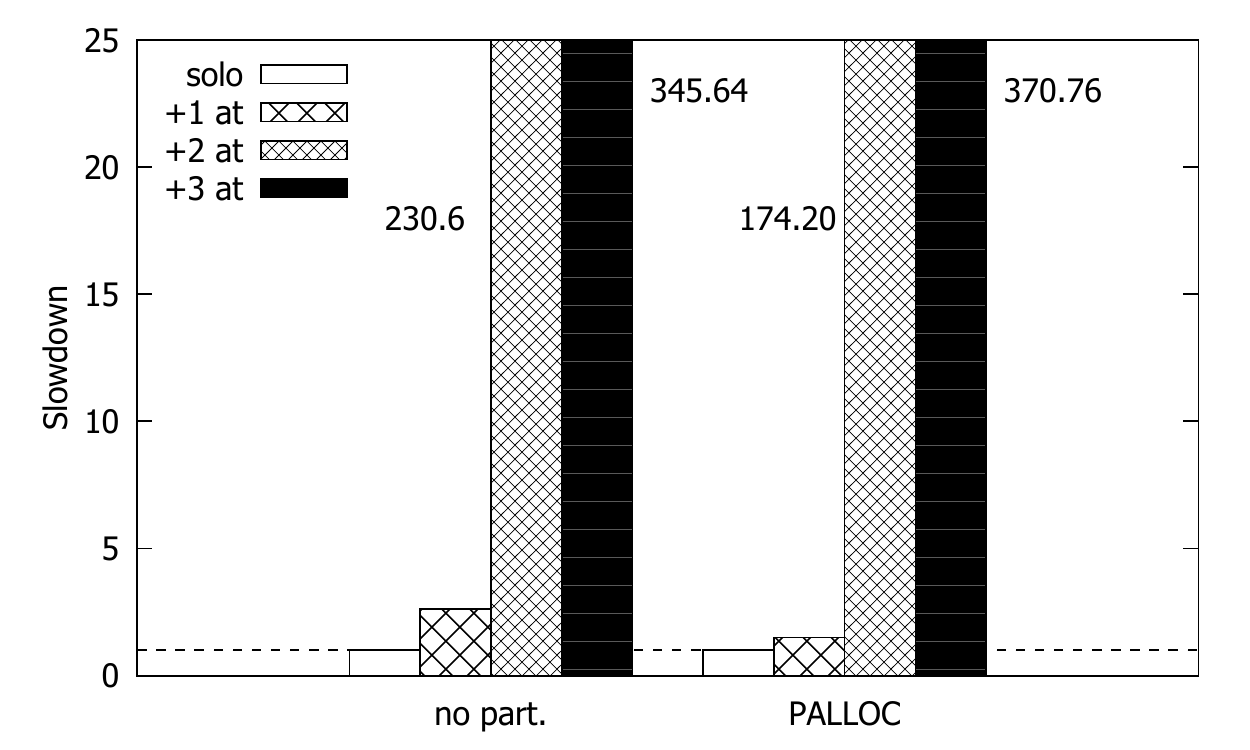}
    \caption{Slowdown }
    \label{fig:bwread-palloc-perf}
  \end{subfigure}
  \begin{subfigure}{0.45\textwidth}
    \includegraphics[width=\textwidth]{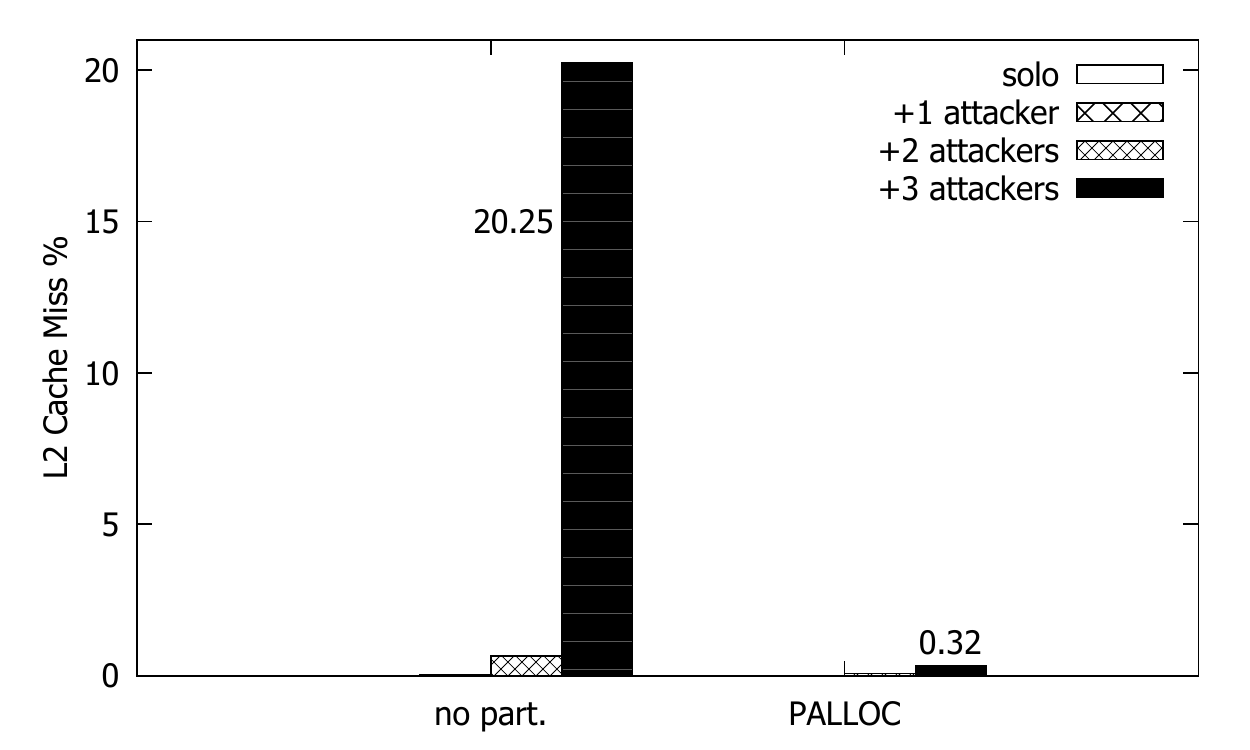}
    \caption{L2 cache miss rate}
    \label{fig:bwread-palloc-misses}
  \end{subfigure}
  \caption{Effect of cache partitioning on Raspberry Pi 3. BwRead
    (LLC) victim vs. BwWrite (DRAM) attackers, w/o and with L2 cache
    partitioning.
  }
  \label{fig:bwread-palloc}
\end{figure}

Figure~\ref{fig:sys_bench_write}, which uses BwWrite (DRAM) attackers
instead, is therefore \emph{surprising} because we observe extreme
performance impacts on two recent in-order architecture based
platforms, the Raspberry Pi 3 and Odroid-C2, both of which feature
four ARM Cortex-A53 cores. In  particular, we observe up to \emph{346X
 slowdown} on the Raspberry Pi 3 platform, which is, to the best of our
knowledge, the highest shared resource contention induced execution
time increase ever reported in literature and it is much worse than
the slowdown observed on the out-of-order Cortex-A15 in Odroid-XU4.
If MSHR contention was the cause of these extreme
performance impacts we observed in the two in-order Cortex-A53
platforms, then we would also expect to see considerable performance
impacts when the read attackers stressed the MSHRs in the previous
experiment, which, however, was not the case.


\subsection{Impact of Cache Partitioning}
As reported in~\cite{valsan2016taming}, we also find that cache
partitioning does not help protect the victim's performance even
when the victim's working-set size fits entirely in its given dedicated
cache partition. Figure~\ref{fig:bwread-palloc} shows the impact of
cache partitioning. For cache partitioning, we use PALLOC~\cite{yun2014rtas},
which implements a page coloring based
kernel-level memory allocator, to equally partition the L2 cache among
the cores in the partitioning setup.
Note that the victim task, BwRead (LLC), suffers
similar degrees of performance impacts regardless of whether partitioning
is applied or not, even while the victim's L2 cache miss rate is
significantly reduced with the cache partitioning---from 20.25\% to
0.32\%---in the presence of three write attackers.
In other words, cache partitioning eliminates unwanted cache-line
evictions from the attackers but it does not help provide cache
performance isolation to the victim, which accesses a dedicated cache
space partition.

\subsection{Impact of Hardware Prefetcher}
Another interesting observation in Figure~\ref{fig:sys_bench_write} is
that although both Cortex-A53 and Cortex-A7 are in-order core designs,
they show dramatically different behaviors in the presence of the
write attackers---Cortex-A53 shows extreme execution time
increases while Cortex-A7 shows no significant execution time increase.

To better understand the root cause(s) of this difference, we compared
cache related performance counter statistics of the Raspberry Pi 3 and
the Odroid C2, both Cortex-A53 based, to the Raspberry Pi 2, which is
based on older Cortex-A7.

\begin{figure}[t]
  \centering
  \begin{subfigure}{0.45\textwidth}
    \includegraphics[width=\textwidth]{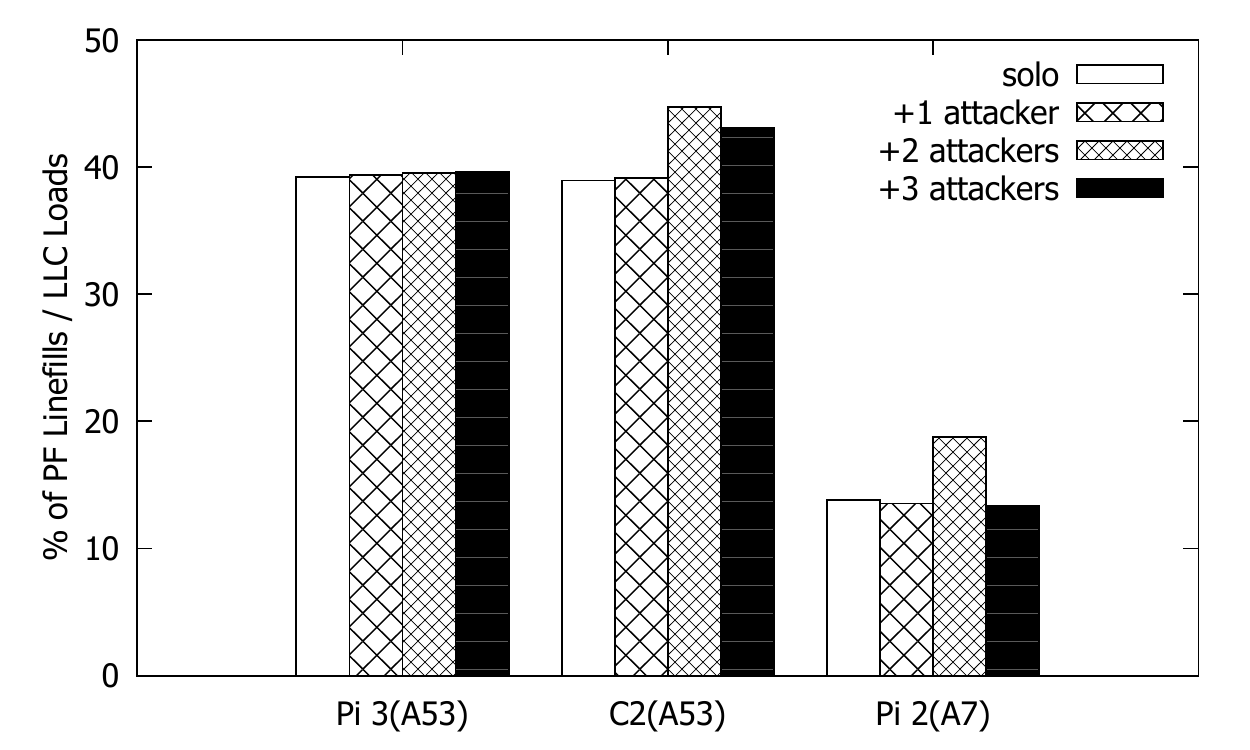}
    \caption{BwRead Co-runners}
    \label{fig:linefill_read}
  \end{subfigure}
  \hfill
  \begin{subfigure}{0.45\textwidth}
    \includegraphics[width=\textwidth]{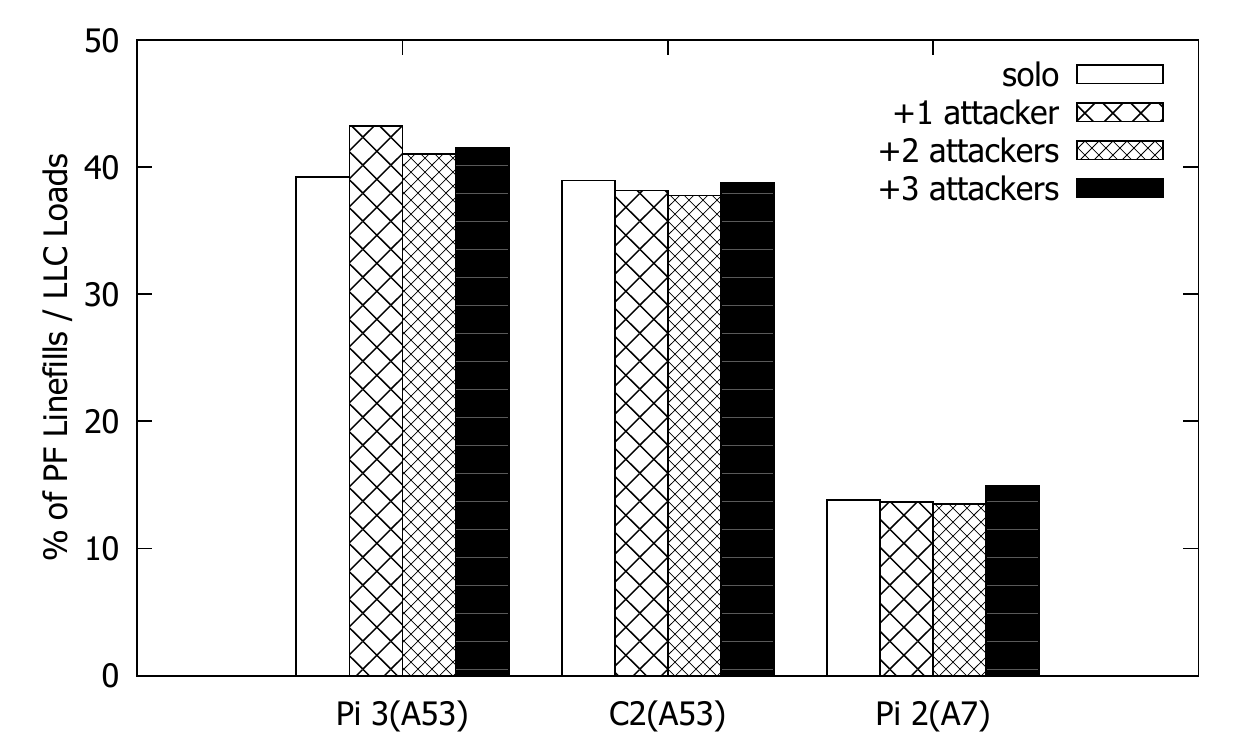}
    \caption{BwWrite Co-runners}
    \label{fig:linefill_write}
  \end{subfigure}
  \caption{Percentage of prefetcher linefills over LLC loads.}
  \label{fig:linefill-ratio}
\end{figure}

The basic experiment setup is the same as above: The BwRead(LLC) victim
task runs on Core 0 in the presence of an increasing number of
BwRead(DRAM) or BwWrite(DRAM) attackers on the rest of the cores.

\begin{figure*}[t]
  \centering
  \begin{subfigure}{0.45\textwidth}
    \includegraphics[width=\textwidth]{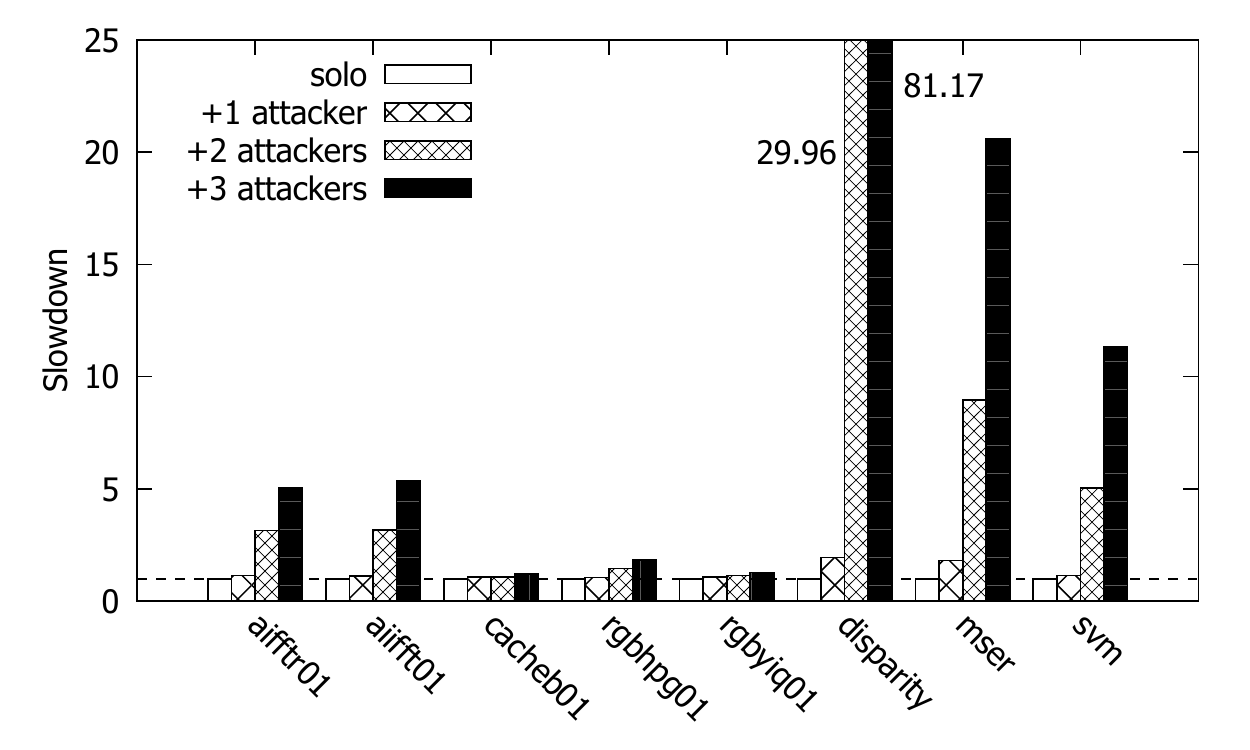}
    \caption{Raspberry Pi 3}
    \label{fig:eembc_pi3}
  \end{subfigure}
  \begin{subfigure}{0.45\textwidth}
    \includegraphics[width=\textwidth]{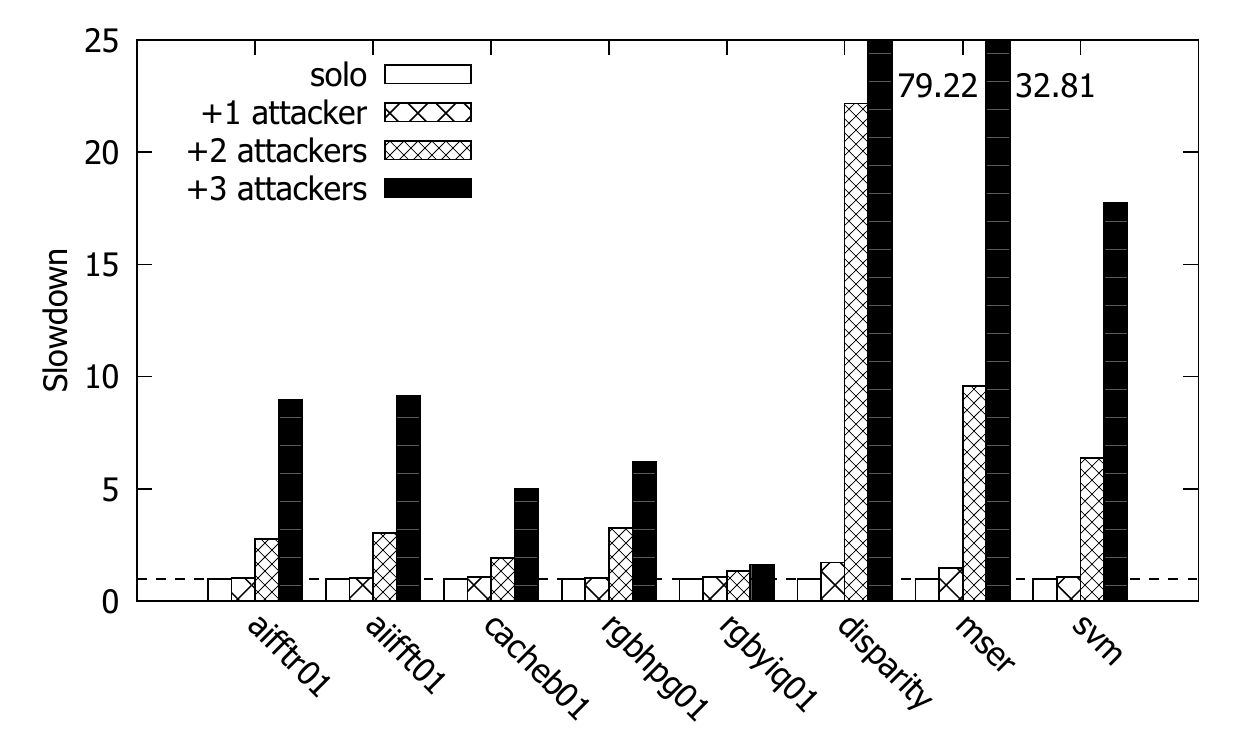}
    \caption{Raspberry Pi 3 (Partitioned shared L2 cache)}
    \label{fig:eembc_pi3_palloc}
  \end{subfigure}
  \begin{subfigure}{0.45\textwidth}
    \includegraphics[width=\textwidth]{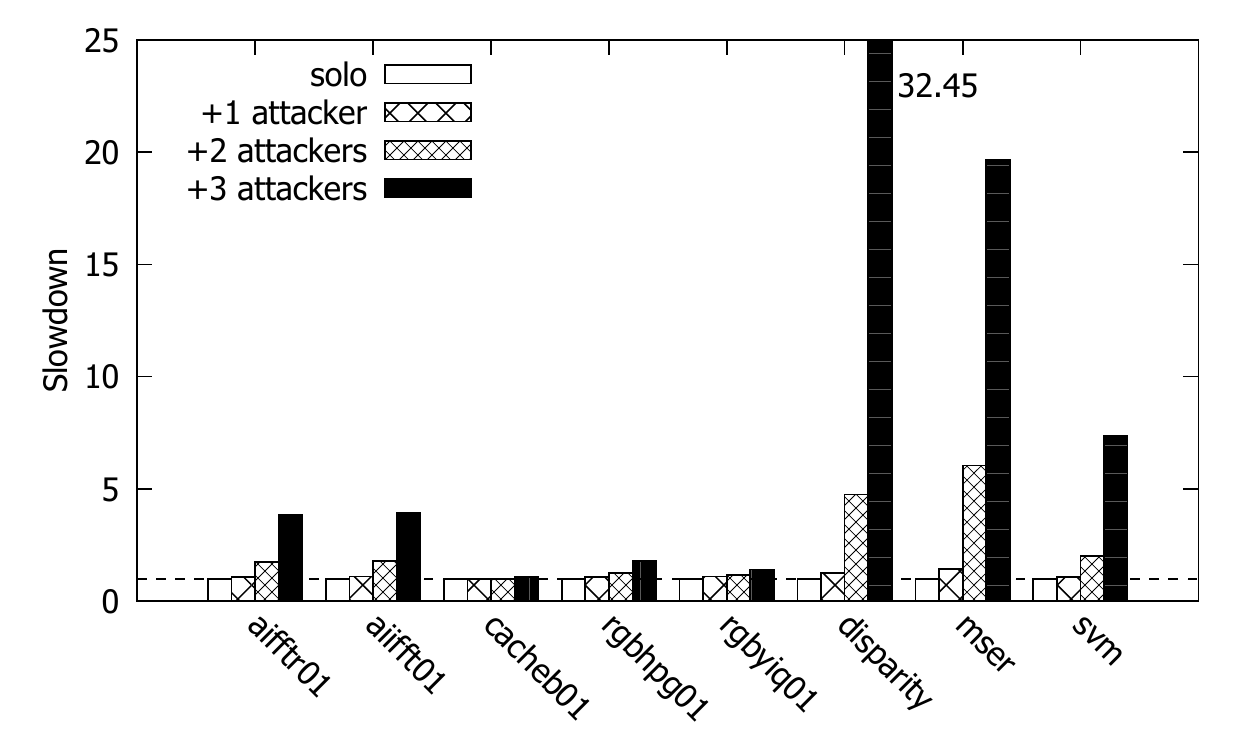}
    \caption{Odroid C2}
    \label{fig:eembc_c2}
  \end{subfigure}
  \begin{subfigure}{0.45\textwidth}
    \includegraphics[width=\textwidth]{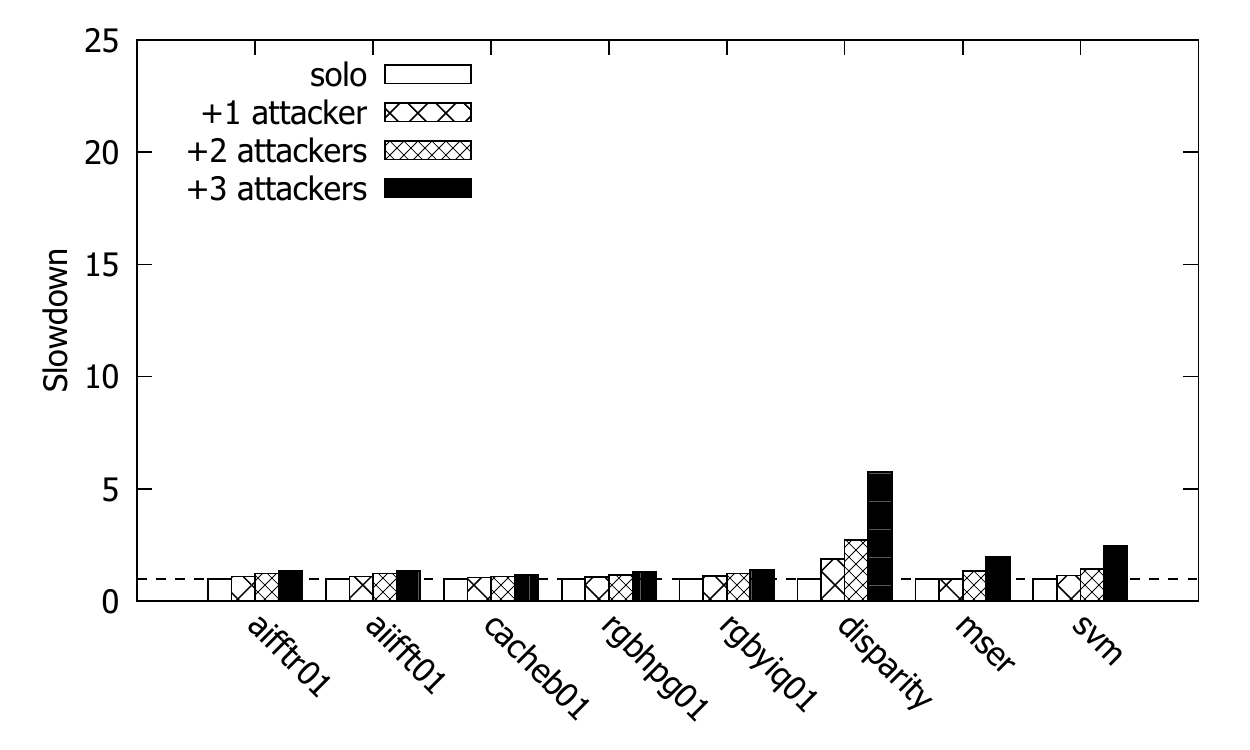}
    \caption{Raspberry Pi 2}
    \label{fig:eembc_pi2}
  \end{subfigure}
  \begin{subfigure}{0.45\textwidth}
    \includegraphics[width=\textwidth]{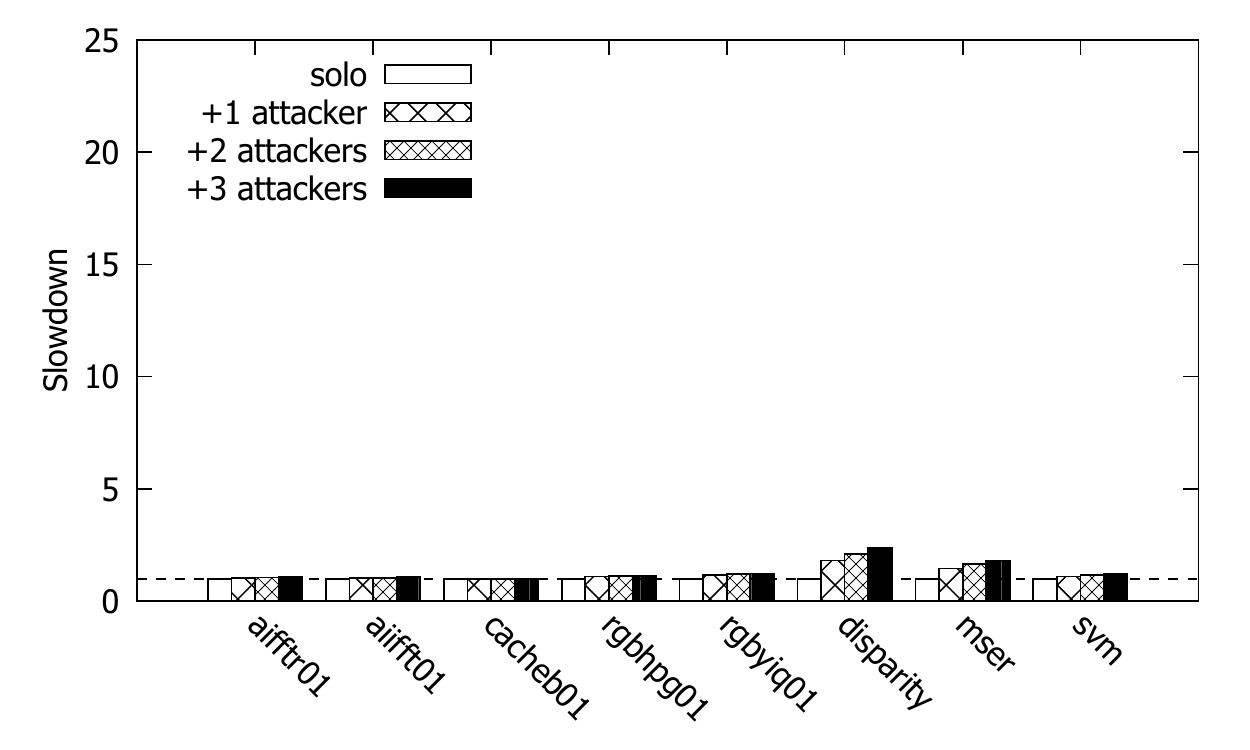}
    \caption{Odroid XU4(A7)}
    \label{fig:eembc_xu4}
  \end{subfigure}
  \begin{subfigure}{0.45\textwidth}
    \includegraphics[width=\textwidth]{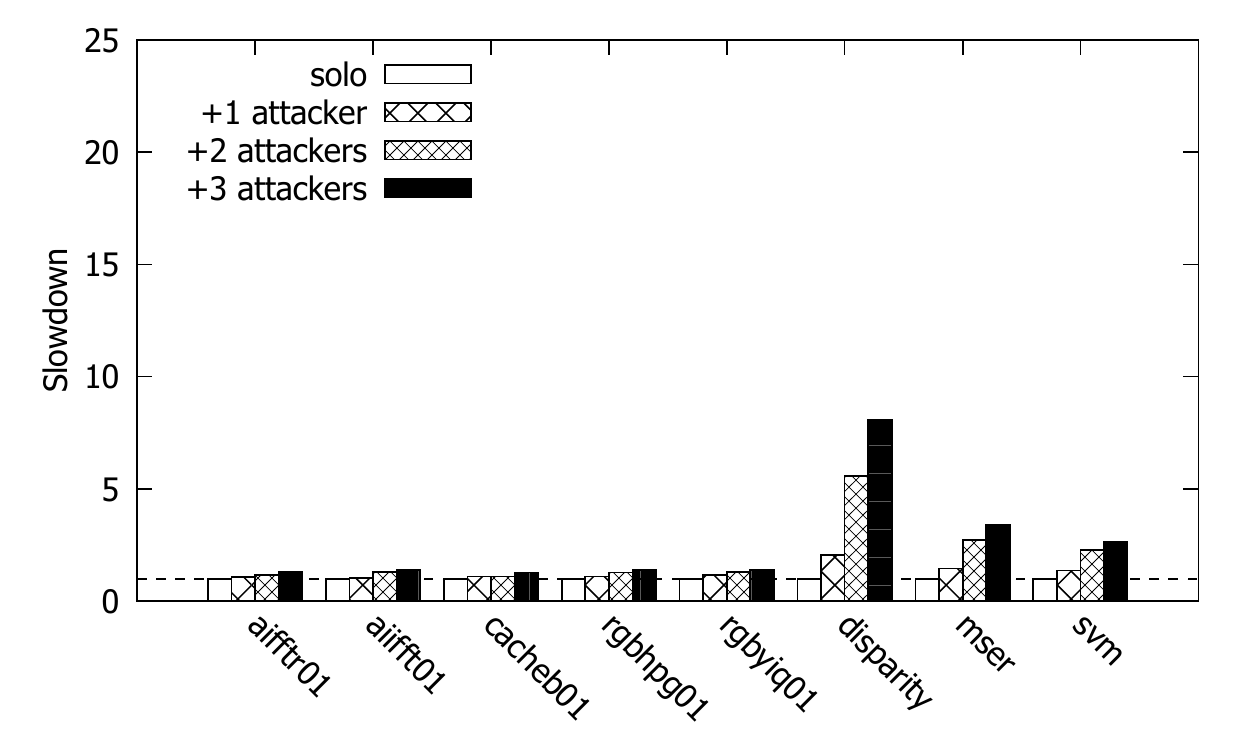}
    \caption{Odroid XU4(A15)}
    \label{fig:eembc_xu4_a15}
  \end{subfigure}
  \caption{EEMBC and SD-VBS benchmarks vs. BwWrite(DRAM) attackers.
    From left to right, \emph{aifftr01, aiifft01, cacheb01, rgbhpg01,
      rgbyiq01} are EEMBC benchmarks and \emph{disparity, mser, svm} are
    SD-VBS benchmarks. For SD-VBS benchmarks, we use \emph{sqcif}
    inputs so that their working-sets can mostly fit in the shared L2
    cache.
  }
  \label{fig:eembc}
\end{figure*}

Our main finding is that all three processors we tested
utilize hardware prefetchers on their L1 data caches, but the
aggressiveness of the hardware prefetchers varies considerably between
the platforms, specifically between the more recent Cortex-A53 and the
older Cortex-A7.

Concretely, Figure~\ref{fig:linefill-ratio} shows the fraction of
the prefetch requests over all LLC memory accesses of the victim task.
Note that the L1 prefetchers on the Pi 3 and C2 generate considerably
more cache-line refills over total L1-D cache-line refills. On both
platforms, the prefetchers account for $\sim$40\% of the total
cache refills, while the Pi 2's prefetcher only accounts for
$\sim$13\%. In other words, the hardware prefetcher of the Cortex-A53
in the Raspberry Pi 3 and Odroid C2 is more aggressive than that of
the Cortex-A7 in Raspberry Pi 2.

Note that a L1 prefetcher's data prefetches may be issued concurrently
with the core's demand request if the L1 data cache itself is a
non-blocking cache. According to the Cortex-A53
documentation~\cite{arm-cortex-a53}, its L1 data cache supports up to
three outstanding cache-misses, suggesting that may indeed be the case.
On the other hand, Cortex-A7's L1 data cache does not appear to support
multiple outstanding cache-misses~\cite{arm-cortex-a7}. Thus, we
believe that Cortex-A7's prefetcher may only be able to prefetch when
the L1 data cache is not being used by the core.
This difference in the L1 data cache's supported concurrent outstanding
misses is important
because cache DoS attacks require concurrent accesses to the shared L2
cache that overflows the cache's internal hardware buffers, namely the
MSHRs and writeback buffer.

\subsection{Impact on Real-World Applications}\label{sec:eval-real}

We also use a set of real-world benchmarks from the
EEMBC~\cite{eembc} and SD-VBS~\cite{venkata2009sd} benchmark suites to
investigate impacts of cache DoS attacks on real-world
applications.
The basic experiment setup is the same as before where we subject each
of the tested benchmarks (the victim) to an increasing number of
BwWrite(DRAM) co-runners (the attackers) on different cores of the
tested multicore platform.

Figure~\ref{fig:eembc} shows the results.
Note, first, that EEMBC
benchmarks generally experience much less performance impact than
SD-VBS benchmarks. This is because most EEMBC benchmarks do not
frequently access the shared L2 cache due to their relatively smaller
working-set sizes (which mostly fit in the L1 data cache), while the SD-VBS
benchmarks access the shared L2 cache much more frequently due to
their larger working-set sizes. Still, on the two Cortex-A53 based
platforms, the Raspberry Pi 3 and Odroid C2, even the EEMBC benchmarks
suffer up to 5.4X and 3.9X slowdown, respectively.
More surprisingly, SD-VBS benchmarks suffer up to 81X slowdown on the
Raspberry Pi 3 and up to 32X slowdown on the Odroid C2. In contrast, the
Odroid-XU4's Cortex-A15 suffers relatively little as it experiences up to 8X
slowdown, which is still significant.

Again, cache partitioning is ineffective
in defending against these cache DoS attacks, as shown in Figure~\ref{fig:eembc_pi3_palloc}.
Instead, cache partitioning was actually detrimental when attacker BwWrite
tasks were present as 7 of the 8 benchmarks suffered worse slowdowns
with partitioning enabled (only the \textit{disparity} benchmark had slightly improved performance).

Lastly, we also evaluate the impacts of BwRead (DRAM) attackers (the
read attackers) on these real-world benchmarks.
Although we do not include here, due to space considerations,
as we observed in experiments using synthetic workloads
(Section~\ref{sec:eval-synthetic}), the read attackers have much less
performance impact (less than 8\% in EEMBC and 2.3X in SD-VBS).

In short, we find that cache DoS attacks, especially the write
attackers targeting cache writeback buffers, are highly effective in
some in-order architecture based embedded multicore platforms, notably
Cortex-A53 based ones. On the other hand, read attackers, which mainly
target cache MSHRs, are not effective on the tested in-order
multicores while they still have considerable timing impacts in
out-of-order architecture based multicore platforms, as suggested
in~\cite{valsan2016taming}.

\begin{table*}[t]
  \centering
  \begin{adjustbox}{width=0.85\textwidth}
  \begin{tabular} {| c | c |}
    \hline
    Core & Quad-core, 1.5 GHz, IQ: 96, ROB: 128, LSQ: 48/48 \\ \hline
    L1-I/D caches & Private 32 kB (2-way), Private 32 kB (4-way),
    MSHRs: 1 (I), 3 (D), Writeback Buffer: 1 (I), 3(D) \\ \hline
    L1-D PF & Stride, Degree: 5, Queue size: 5 \\ \hline
    L2 cache & Shared 512 kB (16-way), MSHRs: 24, Writeback Buffer: 8, hit latency: 12, LRU \\ \hline
    L2 PF & Stride, Degree: 8, Queue size: 8 \\ \hline
    DRAM controller & Read/write buffers: 64, open-adaptive page policy \\ \hline
    DRAM module & DDR3@800MHz, 1 rank, 8 banks\\ \hline
  \end{tabular}
  \end{adjustbox}
  \caption{Baseline simulation parameters for Gem5 and Ramulator.}
  \label{tbl:gem5-setup}
\end{table*}


\section{Understanding Shared Cache Blocking Due to Cache Writeback Buffer}~\label{sec:simulation}

In this section, we study writeback buffer induced shared cache
blocking using a cycle accurate full system simulator.
Specifically, we use Gem5~\cite{binkert2011gem5} and
Ramulator~\cite{kim2016ramulator} to model the CPU and the
memory subsystem, respectively. Table~\ref{tbl:gem5-setup} shows the
baseline configuration we used here.



The simulated CPU we use is comprised of four cores.
Each core has its own private (L1)
instruction and data caches. The data cache (L1-D) is modeled as a
non-blocking cache supporting up to three outstanding cache misses, as
found in the Cortex-A53~\cite{arm-cortex-a53}. The instruction cache (L1-I),
on the other hand, supports up to one outstanding cache-miss.
The instruction cache is then paired with a tagged prefetcher and the
data cache is paired with a stride prefetcher, each of which has its
own internal prefetch queue to hold prefetch addresses before they are
sent to the respective cache.
All cores have access to a single shared L2 cache which has the same
queue structures as the L1 caches and, like the L1-D caches, is paired
with a stride prefetcher. The shared L2 cache is then connected to
a main memory controller (simulated by Ramulator~\cite{kim2016ramulator}).

Note that we carefully configure the prefetchers and L1 data caches
such that cache blocking \emph{cannot} occur due to MSHR contention.
That is, our L2 cache has a sufficient number of MSHRs to support
up to 24 concurrent cache misses, which is enough to support 12
concurrent requests from the cores (their L1 data caches and
prefetchers) and 8 prefetch requests from the L2 prefetcher.
In other words, we removed the possibility of MSHR contention, as
suggested in~\cite{valsan2016taming}. Thus, observed L2 cache
blocking, if any, is not caused by MSHR exhaustion, but instead by
writeback buffer exhaustion of the L2 cache.

\subsection{Effect of Hardware Prefetchers}

In this experiment, we investigate the impacts of L1 and L2
prefetchers on the effectiveness of cache DoS attacks, targeting the
cache writeback buffer. The experiment setup is the same as
before---specifically, we run the BwRead (LLC) victim and three
BwWrite (DRAM) attackers. We repeat the experiment in different
L1-D and L2 prefetcher configurations.

\begin{figure}[h]
  \centering
  \includegraphics[width=.45\textwidth]{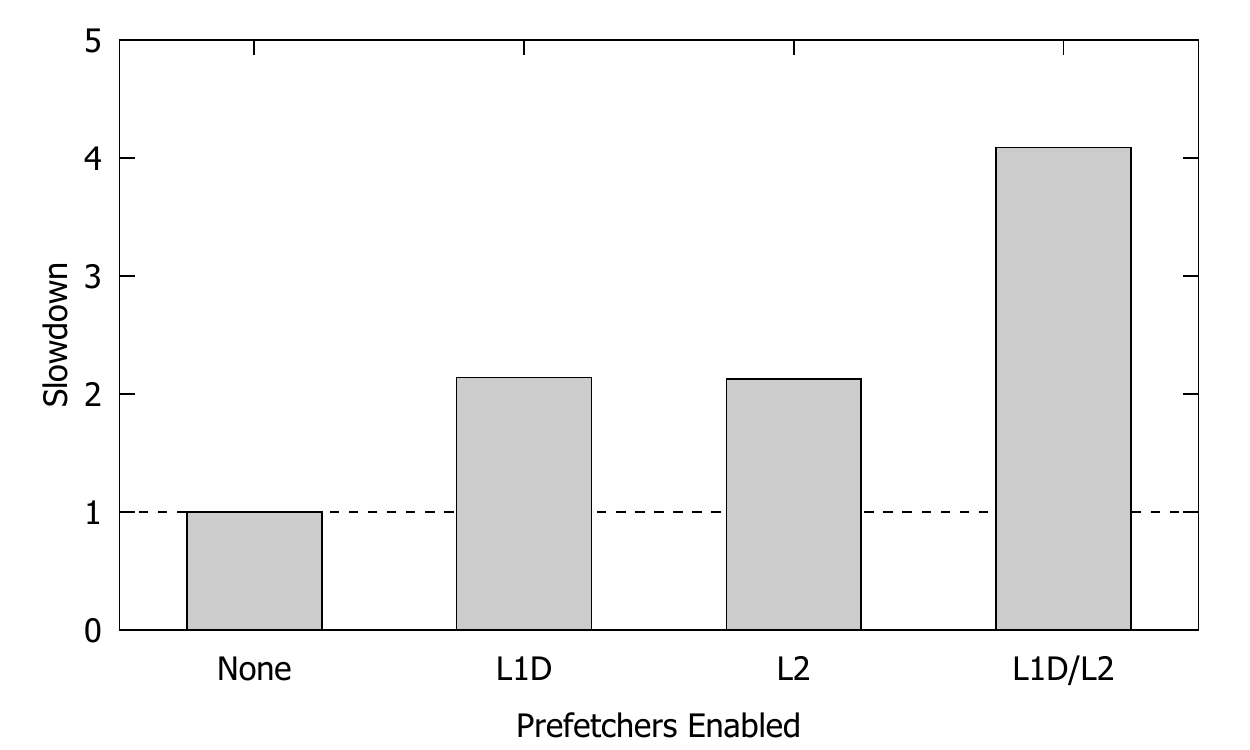}
  \caption{Effects of hardware prefetchers.}
  \label{fig:enable-prefetchers}
\end{figure}

Figure~\ref{fig:enable-prefetchers} shows the results.
When the L1-D or L2 prefetchers are enabled (labeled `L1D' and `L2'), the
performance of the victim task becomes noticeably worse as we observe
more than 2X execution time increases. When we enable both L1-D and L2
prefetchers (labeled `L1D/L2'), the result is more than 4X execution time
increase, compared to the  configuration where all prefetchers are
disabled (labeled `None').

In other words, enabling hardware prefetchers
increases the victim's execution time due to increased L2 cache
blocking, driven by increased cache writebacks initiated by the
additional prefetch refill requests at the L2 cache.


\subsection{Effect of Writeback Buffer Size}

In this experiment, we explore the impacts of writeback buffer size of
the shared L2 cache to the effectiveness of cache DoS attacks,
targeting the writeback buffer.
Specifically, we want to know if increasing the size of the L2 writeback
buffer reduces L2 cache blocking, which in turn would improve the victim
task's performance.

\begin{figure}[h]
  \centering
  \includegraphics[width=.45\textwidth]{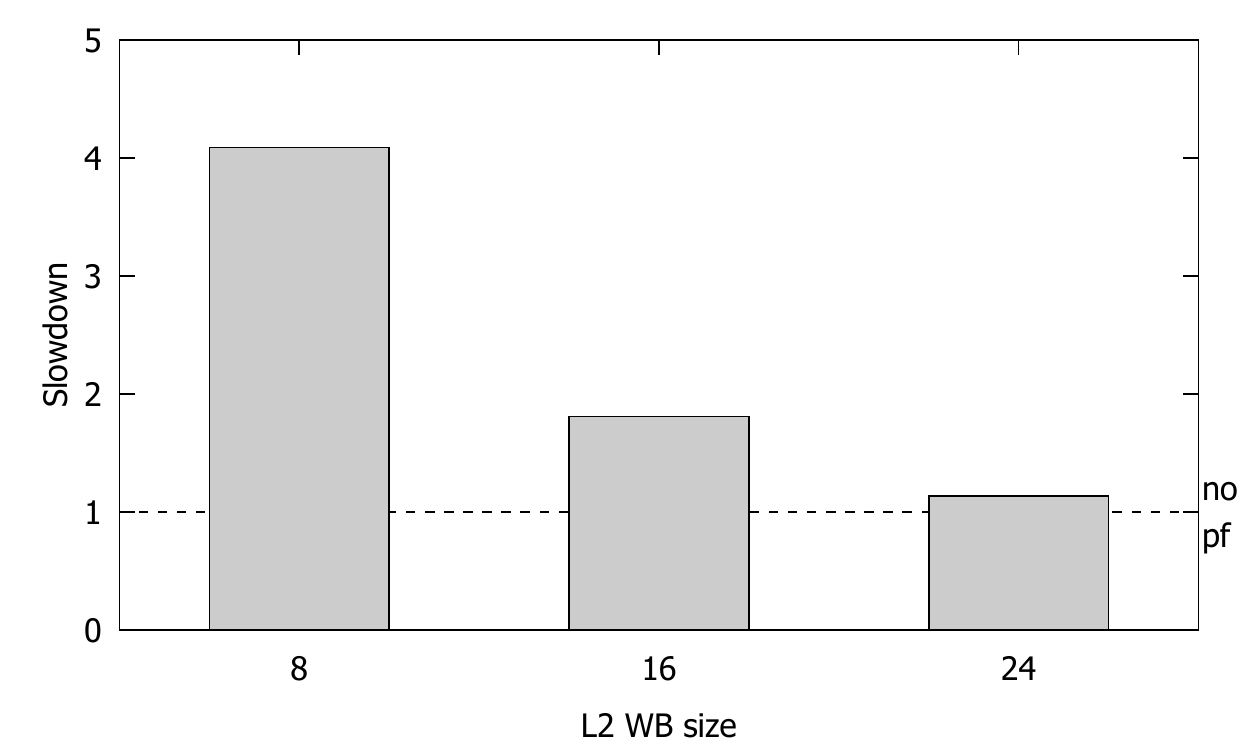}
  \caption{Effect of the L2 cache writeback buffer size.}
  \label{fig:buffer-size}
\end{figure}

\begin{figure}[h]
  \centering
  \includegraphics[width=.45\textwidth]{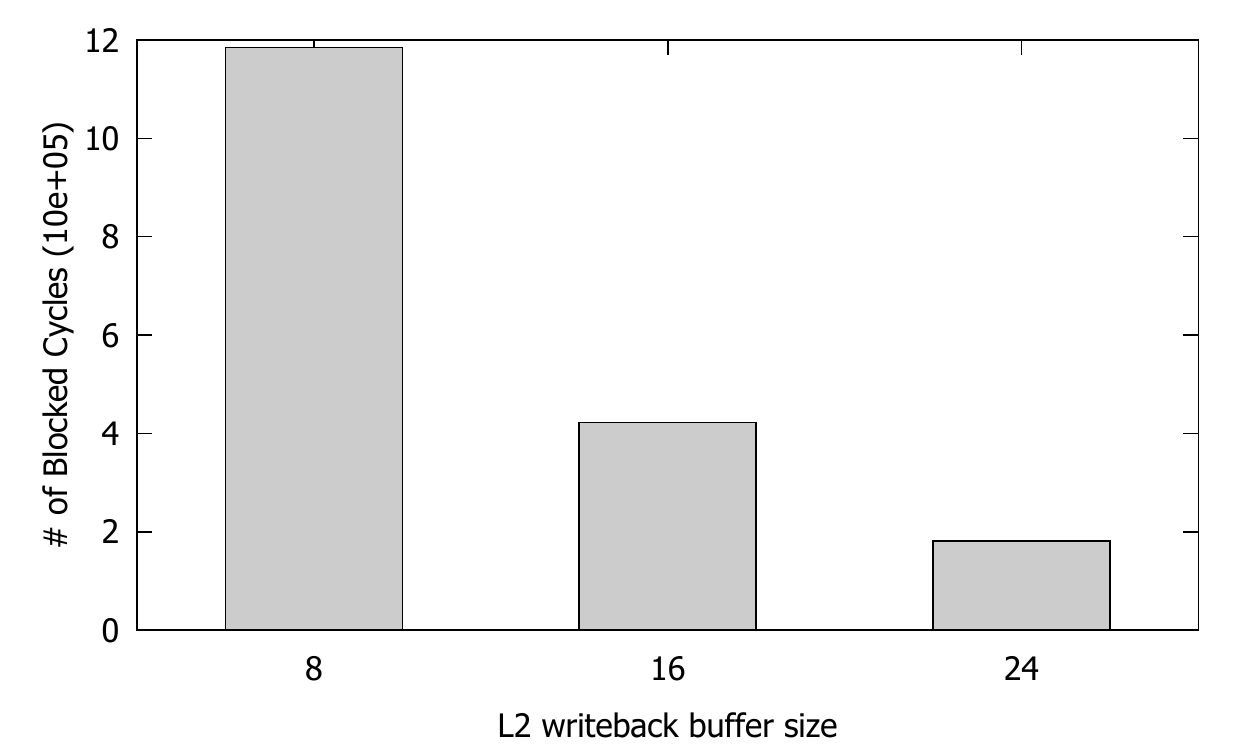}
  \caption{Blocked cycles vs. the L2 writeback buffer size.}
  \label{fig:blocked-cycles-L2}
\end{figure}

Figure~\ref{fig:buffer-size} shows the results.
As suspected, when we increase the size of the L2 writeback buffer, the
performance of the victim task is improved accordingly. This is
because the L2 cache's blocked time is decreased.
Figure~\ref{fig:blocked-cycles-L2} shows the total number of cycles
during which the L2 cache is blocked in the same experiment.

In summary, we find that the presence of hardware prefetchers and the
size of the L2 cache writeback buffer are major factors affecting the
platform's susceptibility to cache DoS attacks.


\section{OS-level Defense Mechanism Against Cache DoS Attacks} \label{sec:solutions}

In this section, we present an OS-level solution to prevent
denial-of-service attacks on the shared cache in a multicore platform, especially
those targeting the cache writeback buffer.


Our solution is software-based and is built on top of an existing
memory bandwidth throttling mechanism called
MemGuard~\cite{yun2013rtas}. MemGuard uses per-core hardware performance
counters to regulate (throttle) each core's maximum memory bandwidth
usage. Specifically, it uses the \emph{LLC miss} counter to calculate
the amount of memory bandwidth consumed by each core. Prior
studies show the effectiveness of memory bandwidth throttling in
protecting real-time tasks
~\cite{nowotsch2014multi,pellizzoni2016memory,bechtel2018picar,agrawal2018analysis}.

However, we find a significant limitation of using the LLC miss count
as a sole means to measure and regulate memory bandwidth because it
effectively treats both read and write misses as equal despite the
fact that write misses may incur additional writeback
traffic on a write-back cache.
While the cache writebacks are typically not in the
critical path and are processed opportunistically in both cache and DRAM
controllers, as discussed in Section~\ref{sec:simulation},
write-backs can block the cache when the writeback
buffer is full. It can also delay cache-line refill operations if the memory
controller cannot process backlogged DRAM writes in the
background~\cite{yun2015ecrts}. Thus, as shown in
Section~\ref{sec:evaluation}, we find that write intensive attackers
are far more impactful than read intensive ones.

To address this limitation, we propose to extend MemGuard by utilizing
an additional performance counter that measures the number of
\textit{LLC writebacks} in addition to the existing counter that monitors
the LLC misses. By using the two counters, we can regulate both the
number of LLC misses and writebacks separately. For example, we can
throttle write intensive tasks more without affecting read intensive
tasks by setting a low threshold for the writeback counter while
setting a high threshold for the cache miss counter.

To demonstrate the effectiveness of this solution, we consider two
different application scenarios.

In the first scenario, we investigate the impact of memory bandwidth
throttling to application performance on the throttled core.
First, we compare the baseline MemGuard and our modified version by
applying them to throttle BwRead (DRAM) and BwWrite (DRAM) subject
tasks in isolation (i.e., one task at a time). For the baseline
MemGuard, we set the LLC miss threshold (read memory bandwidth) to 100
MB/s, and for our  modified MemGuard, we set the LLC miss threshold (read)
to 500 MB/s, while additionally setting the LLC writeback threshold
(write) to 100 MB/s.

\begin{figure}[t]
  \centering
  \includegraphics[width=0.45\textwidth]{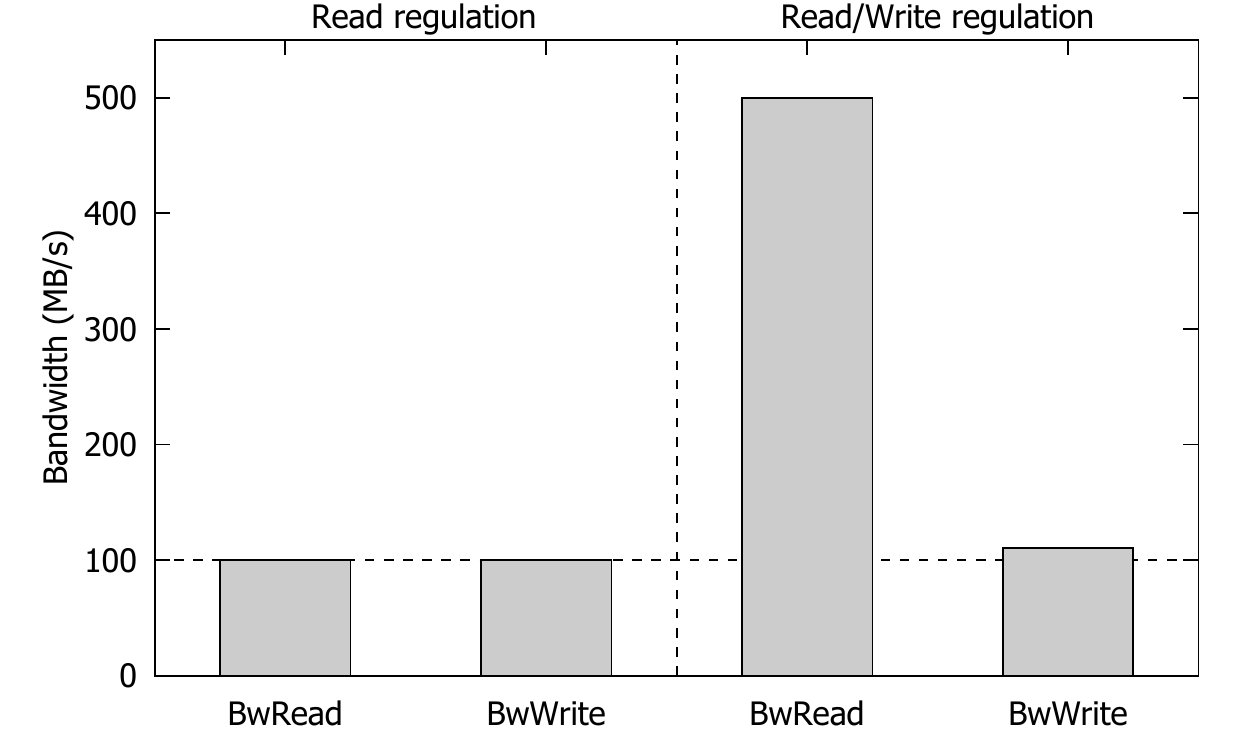}
  \caption{ Effect of read-only (MemGuard~\cite{yun2013rtas}) vs.
    separate read/write bandwidth regulation (Our approach).
  }
  \label{fig:new-memguard}
\end{figure}

\begin{figure}[h]
  \centering
  \includegraphics[width=0.45\textwidth]{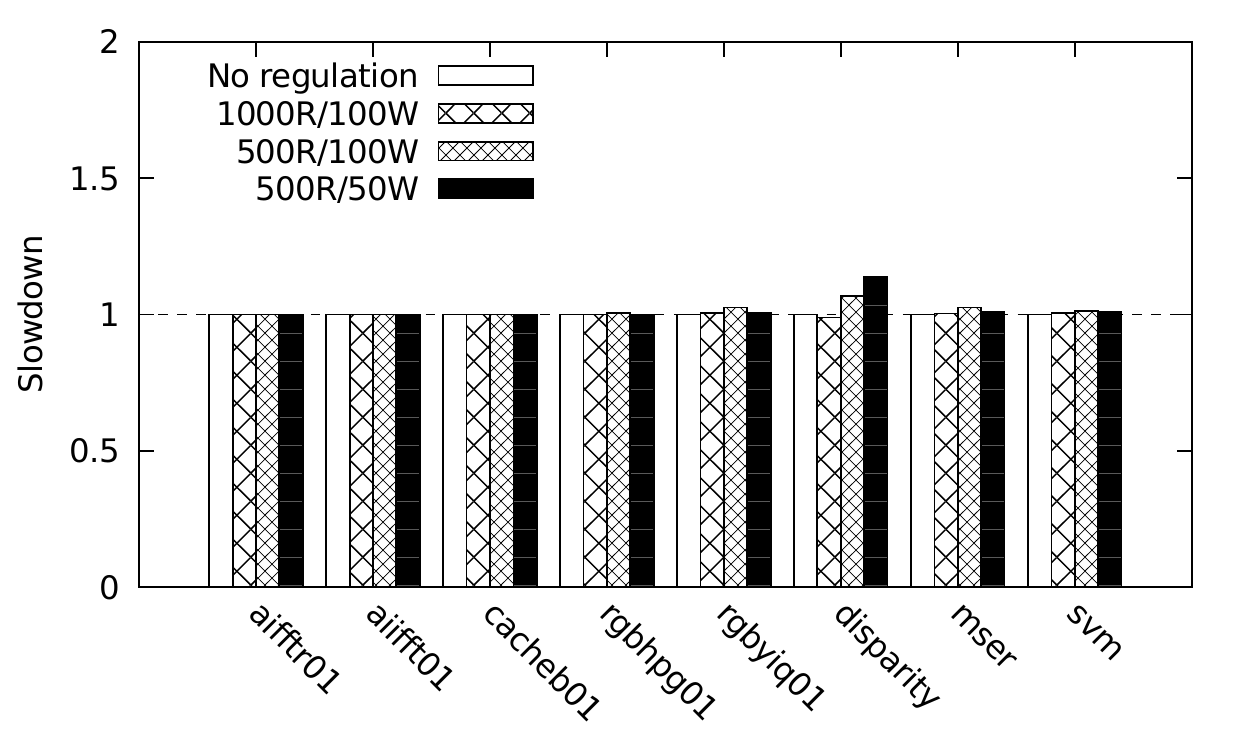}
  \caption{Effect of read/write bandwidth regulation on the regulated
    non-attacker tasks.}
  \label{fig:baseline}
\end{figure}

\begin{figure}[t]
  \centering
  \includegraphics[width=0.45\textwidth]{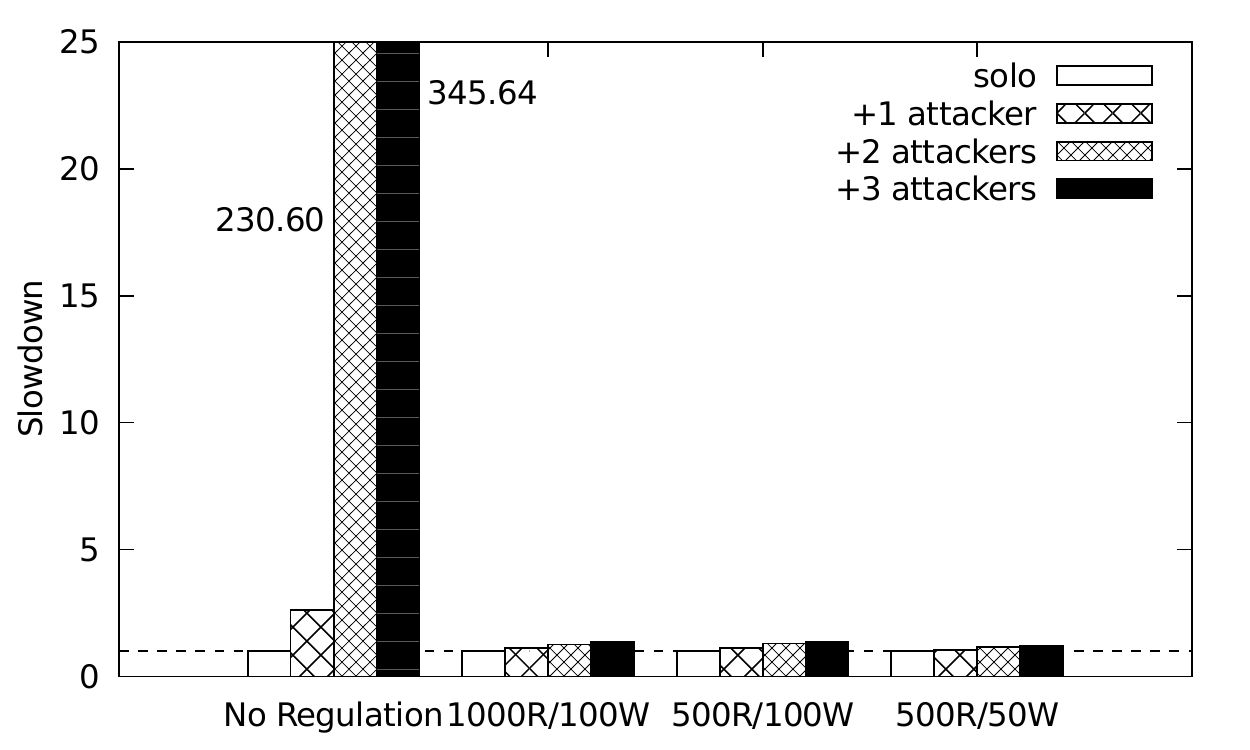}
  \caption{Effect of read/write bandwidth regulation on protecting
    BwRead (LLC) victim from write DoS attacks.}
  \label{fig:memguard_solution}
\end{figure}

\begin{figure}[t]
  \centering
  \includegraphics[width=0.45\textwidth]{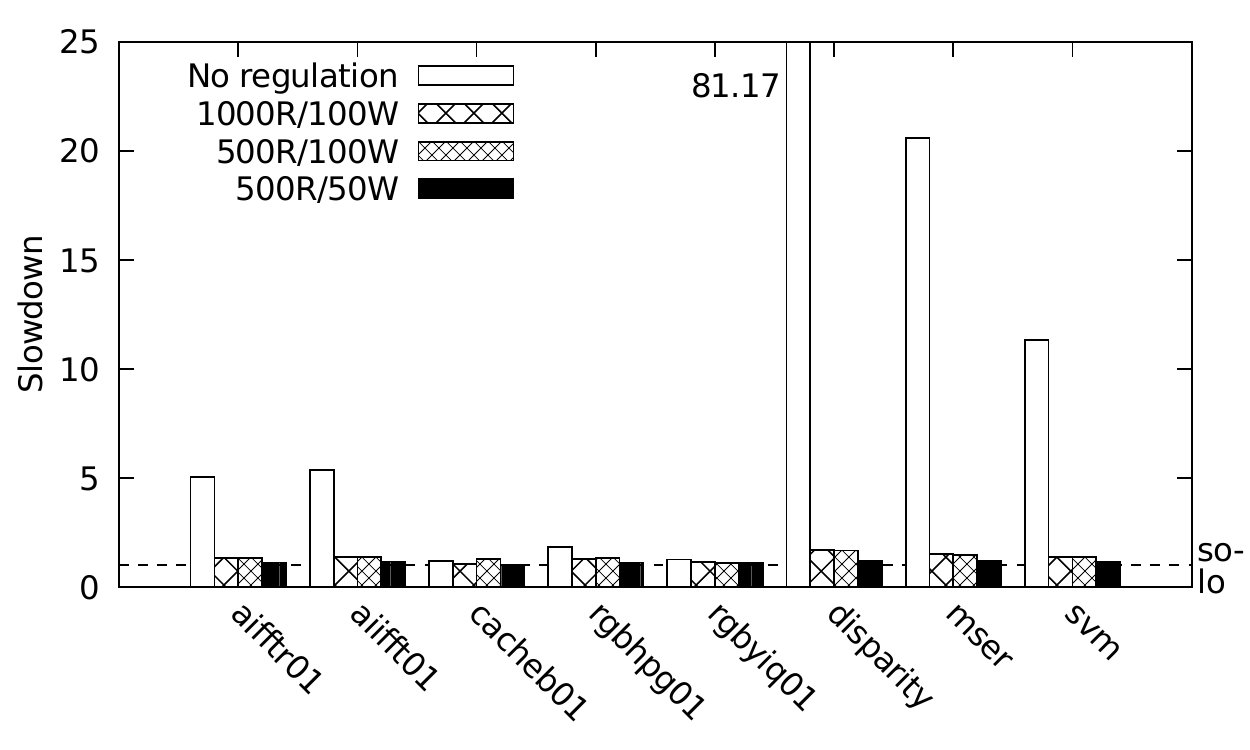}
  \caption{Effect of read/write bandwidth regulation on protecting EEMBC and
    SD-VBS victim tasks from write DoS attacks. }
  \label{fig:memguard_solution_rwall}
\end{figure}

Figure~\ref{fig:new-memguard} shows the results. In the baseline
MemGuard, both BwRead and BwWrite tasks are limited to 100 MB/s as
both read and write misses are treated equally.
With our modification, however, the BwWrite task is limited to 100
MB/s, while the BwRead task's performance is increased to 500 MB/s, as
we expected. This means that we can provide the same degree of
interference protection by heavily throttling write memory bandwidth,
while allowing higher read memory bandwidth to the tasks running on
the regulated cores.

In the next experiment,
we use EEMBC and SD-VBS benchmarks and evaluate their performance
impacts under the following three read/write throttling
configurations: \emph{1000R/100W}, \emph{500R/100W}, and
\emph{500R/50W}. In 1000R/100W, we set 1000 MB/s threshold for LLC
misses and 100 MB/s threshold for LLC writebacks. The other two
configurations, 500R/100W and 500R/50W, are similarly defined.

Figure~\ref{fig:baseline} shows the results. Note, first, that for
most benchmarks, read/write throttling does not have any noticeable
performance impact. At 1000R/100W, in particular, the performance impact
of throttling is negligible. This is because the tested benchmarks are
mostly not memory intensive and thus do not exceed the assigned
throttling parameters. As we assign less read and write bandwidth, to
500R/100W and 500R/50W, some benchmarks, notably \emph{disparity},
show performance impacts due to throttling, but the effects are
relatively minor. Note that the effect of throttling may vary
significantly depending on the application characteristics. In
general, memory intensive applications, especially write intensive
ones, will suffer the most, while read intensive ones will suffer much
less under our proposed read/write regulation scheme. Fortunately,
real-world applications are usually more dependent on read bandwidth
rather than write bandwidth, which makes our approach attractive.



In the second scenario, we applied our modified MemGuard to protect
against the write attackers on the Raspberry Pi 3 platform, which
suffered the most severe interference (Section~\ref{sec:evaluation}).
We repeat the experiments in Section~\ref{sec:eval-synthetic} and
Section~\ref{sec:eval-real} with three different read/write throttling
configurations to regulate the write attackers.

Figure~\ref{fig:memguard_solution} shows the results for the BwRead
(LLC) victim task (c.f., Figure~\ref{fig:sys_bench_write} in
Section~\ref{sec:eval-synthetic}). As can be seen, applying
write regulation on the attacker cores is effective in protecting the
victim task, and decreasing the threshold can further improve
performance. Concretely, the victim task's execution time increase is
reduced from 345X down to 1.34X with a write threshold of 100MB/s and
1.2X with a write threshold of 50MB/s. In this experiment, increasing
read regulation from 500MB/s to 1000MB/s has no effect because the
attackers are throttled by the write regulation.

Figure~\ref{fig:memguard_solution_rwall} shows the results for the EEMBC
and SD-VBS benchmark victim tasks in the presence of three write
attackers (c.f., Figure~\ref{fig:eembc_pi3}).
Similar to the BwRead (LLC) experiment above, when write regulation is
applied on attacker cores, the performance of the victim tasks
improve. This is most noticeable in the SD-VBS benchmarks, such as the
disparity benchmark whose WCET increase is reduced from 81.17X to
1.20X with the 500R/50W setup, showing the effectiveness of our
read/write regulation approach in protecting against cache DoS attacks.


In summary, we show that our extended OS-level mechanism, which
regulates read and write bandwidth separately, allows us to apply more
efficient and targeted bandwidth regulation policies that can prevent
cache DoS attacks, while minimizing the performance impact for the
regulated cores with minimal performance impact to non-attacker tasks.

\section{Related Work} \label{sec:related}

In a real-time system, the ability to guarantee predictable
timing is highly important. However, it is difficult to achieve
predictable timing in a multicore platform due to shared hardware
resources, such as cache and main memory.
Therefore, much research effort has been focused on analyzing and
controlling the timing impacts of these shared
hardware resources in multicore platforms.
For shared caches, most prior works focused on cache space partitioning
by using various OS or hardware mechanisms~\cite{kim2017attacking,ward2013ecrts,kim2013coordinated,mancuso2013rtas,kessler1992page,wolfe1994software,liedtke1997controlled,suh2002new,kim2004fair,chandra2005predicting}.
Recently, however, Valsan et al. experimentally showed that cache
space partitioning does not necessarily guarantee cache performance
isolation in non-blocking caches used in modern multicore
processors~\cite{valsan2016taming}. The authors then identified
miss-status-holding-registers (MSHRs), which are the cache's internal
buffers to track outstanding cache-misses, as the source of the
observed timing increases in a number of out-of-order multicore
platforms. In essence, they identified cache MSHRs as a
denial-of-service (DoS) attack vector.
In contrast, our work shows that even relatively simple in-order
multicore platforms are not immune to cache DoS attacks and identifies
an additional internal hardware structure of a non-blocking cache,
namely the writeback buffer, as another DoS attack vector.

Several prior studies investigated various DoS attack
vectors in multicore. Moscibroda et al. examined DoS attacks on
memory (DRAM) controllers~\cite{moscibroda2007memory}. They found that
the commonly used FR-FCFS~\cite{rixner2000memory} scheduling
algorithm, which may re-order memory requests to maximize throughput,
is vulnerable to DoS attacks. They suggested ``fair'' memory
scheduling as a solution. Many subsequent papers proposed various fair
memory scheduling algorithms in DRAM
controllers~\cite{mutlu2007stall,mutlu2008parallelism,kim2010thread,subramanian2013mise}.
Keramidas et al. studied DoS attacks on cache space and proposed a cache
replacement policy that allocates less space to such attackers (or
cache ``hungry'' threads)~\cite{keramidas2006preventing}.
Woo et al. investigated DoS attacks on cache bus (between L1 and L2)
bandwidth, main memory bus (front-side bus) bandwidth, and shared cache
space, on a simulated multicore platform~\cite{woo2007analyzing}.
In contrast, our work demonstrates the feasibility of shared cache DoS
attacks on real multicore platforms and identifies an internal
hardware structure of non-blocking caches as a DoS attack
vector. Furthermore, we present an OS-based solution
on a real multicore platform that prevents identified cache DoS attacks.

Recently, microarchitectural timing attacks\cite{ge2016survey} have gained
significant attention, both from the public and the research community,
in the wake of the Meltdown, Spectre, and Foreshadow
attacks\cite{lipp2018meltdown,kocher2018spectre,vanBulck2018foreshadow}. In
general, these timing attacks aim to gain secret information through
externally observable timing differences in accessing
microarchitectural resources such as cache.
In contrast, DoS attacks on microarchitectural resources, which we
focus on in this paper, aim to directly influence performance (timing)
of the victim applications or cores.
The Rowhammer attack\cite{kim2014flipping} is another kind of attack
targeting hardware. It exploits a reliability failure mode in modern
DRAM hardware where repeatedly and quickly accessing certain DRAM
locations can result in bit flips in nearby memory
locations. Successful attacks that break OS memory isolation boundaries
have been demonstrated in servers and mobile
devices~\cite{seaborn2015exploiting,veen2016drammer}. As
safety-critical embedded real-time systems are becoming more
connected, such as we are already seeing in cars, we believe that software
attacks targeting computer hardware are increasingly important areas
that need attention from both the real-time and security research
communities.

\section{Conclusion}\label{sec:conclusion}

In this paper, we investigated the feasibility and severity of DoS
attacks on shared caches in multicore platforms.
From careful experiments on a number of contemporary embedded
multicore platforms, we observed surprisingly severe execution time
increases---up to 346X slowdown---on some of the tested platforms. In
particular, we found that two recent in-order architecture, ARM
Cortex-A53, based multicore platforms are especially more susceptible
to write-intensive cache DoS attacks than more complex out-of-order
architecture based multicore platforms. 

From detailed micro-architectural analysis using a cycle-accurate full
system simulator, we identified the shared cache's writeback buffer as
a possible DoS attack vector, in addition to the previously known cache
MSHR, that could have caused the behaviors we observed on the real
platforms.

We propose a software (OS) solution to mitigate cache DoS
attacks targeting the shared cache's writeback buffer.
Our solution implements a separate read and write memory
bandwidth regulation mechanism to effectively counter write intensive
cache DoS attacks while minimizing performance impacts to read heavy
normal applications.
Our solution is implemented in Linux and shown to be effective
to counter the shared cache DoS attacks.

As future work, we plan to investigate OS and architecture support
mechanisms to withstand different types of micro-architectural attacks
(e.g., cache timing attacks~\cite{lipp2018meltdown,kocher2018spectre},
Rowhammer attacks~\cite{kim2014flipping}) in the context of
safety-critical real-time systems.

\section*{Acknowledgements} \label{acknowledge}
This research is supported by NSF CNS 1718880, CNS 1815959, and NSA
Science of Security initiative contract \#H98230-18-D-0009.

\bibliographystyle{abbrv}
\bibliography{reference}

\end{document}